\documentclass[12pt]{article}

\def\lromn#1{\uppercase\expandafter{\romannumeral#1}}

\usepackage{graphicx,amsmath,amssymb}
\def\lromn#1{\uppercase\expandafter{\romannumeral#1}}

\usepackage{epsfig}
\usepackage{array}
\usepackage{color}
\usepackage{float}
\usepackage{lscape,graphicx}
\usepackage{graphics}
\usepackage{amssymb}
\def\ltap{\ \raisebox{-.4ex}{\rlap{$\sim$}} \raisebox{.4ex}{$<$}\ }
\def\gtap{\ \raisebox{-.4ex}{\rlap{$\sim$}} \raisebox{.4ex}{$>$}\ }

\def\be{\begin{equation}}
\def\ee{\end{equation}}
\def\gs{\mathrel{
   \rlap{\raise 0.511ex \hbox{$>$}}{\lower 0.511ex \hbox{$\sim$}}}}
\def\ls{\mathrel{
   \rlap{\raise 0.511ex \hbox{$<$}}{\lower 0.511ex \hbox{$\sim$}}}}
\newcommand{\bea}{\begin{equation}\begin{array}{c}}
\newcommand{\eea}{\end{array}\end{equation}}
\newcommand{\ea}{\end{array}}

\newcommand{\beq}{\begin{equation}}
\newcommand{\eeq}{\end{equation}}
\newcommand{\bad}{\begin{array}{ccc}}

\newcommand{\betabeta}{\mbox{$(\beta \beta)_{0 \nu}  $}}

\begin{document}

\begin{flushright}
\today \\
\end{flushright}

\hfill{{\small Ref. SISSA 26/2012/EP}}

\hfill{{\small Ref. OU-HET 756/2012}}

\begin{center}
\mathversion{bold}
\vspace{0.2cm}
\mathversion{bold}
{\bf{\large Observables in Neutrino Mass Spectroscopy Using Atoms }}\\

\mathversion{normal}

\vspace{0.4cm}
D. N. Dinh$\mbox{}^{a,b)}$,
S. T. Petcov$\mbox{}^{a,c)}$
\footnote{Also at: Institute of Nuclear Research and
Nuclear Energy, Bulgarian Academy of Sciences, 1784 Sofia, Bulgaria}
N. Sasao$\mbox{}^{d)}$,
M. Tanaka$\mbox{}^{e)}$ and
M. Yoshimura$\mbox{}^{f)}$

\vspace{0.2cm}
$\mbox{}^{a)}${\em  SISSA and INFN-Sezione di Trieste, \\
Via Bonomea 265, 34136 Trieste, Italy.\\}

\vspace{0.1cm}
$\mbox{}^{b)}${\em Institute of Physics, Vietnam Academy of Science and Technology, \\ 10 Dao Tan, Hanoi, Vietnam.\\
}

\vspace{0.1cm}
$\mbox{}^{c)}${\em Kavli IPMU, University of Tokyo (WPI), Tokyo, Japan.\\}

\vspace{0.1cm}
$\mbox{}^{d)}${\em Research Core for Extreme Quantum World,
Okayama University,\\ Okayama 700-8530 Japan\\}

\vspace{0.1cm}
$\mbox{}^{e)}${\em Department of Physics, Graduate School of Science,
Osaka University,\\
Toyonaka, Osaka 560-0043, Japan.\\}

\vspace{0.1cm}
$\mbox{}^{f)}${\em Center of Quantum Universe, Faculty of Science,
Okayama University,\\
Okayama 700-8530, Japan.\\}

\end{center}

\vspace{1cm}

\begin{center}
\begin{Large}
{\bf Abstract}
\end{Large}
\end{center}
{
 The process of collective de-excitation of
atoms in a metastable level into emission
mode of a single photon plus a neutrino pair,
called radiative emission of neutrino pair
(RENP), is sensitive to the
absolute neutrino mass scale,
to the neutrino mass hierarchy and
to the nature  (Dirac or Majorana) of
massive neutrinos.
We investigate how the indicated neutrino
mass and mixing observables can be determined
from the measurement of the
corresponding continuous photon spectrum taking
the example of a transition between
specific levels of the Yb atom.
The possibility of determining the nature of
massive neutrinos and, if neutrinos are Majorana fermions,
of obtaining information about the Majorana phases in
the neutrino mixing matrix, is analyzed in the cases
of normal hierarchical, inverted hierarchical and
quasi-degenerate types of neutrino mass spectrum.
We find, in particular, that the sensitivity to the nature of
massive neutrinos depends critically
on the atomic level energy difference
relevant in the RENP.
}
\newpage

\section{Introduction}
  Determining the absolute scale of neutrino masses,
the type of neutrino mass spectrum, which can be either
with normal or inverted ordering
\footnote{We use the convention adopted in \cite{PDG12}.}
(NO or IO),
the nature (Dirac or Majorana) of massive neutrinos,
and getting information about the Dirac and Majorana
CP violation phases in the neutrino mixing matrix,
are the most pressing and challenging problems
of the future research in the field of
neutrino physics (see, e.g., \cite{PDG12}).
 At present we have compelling evidence for
existence of mixing of three massive neutrinos
$\nu_i$, $i=1,2,3$, in the weak charged lepton current
(see, e.g., \cite{Nu2012}). The masses $m_i \geq 0$ of the three
light neutrinos $\nu_i$ do not exceed a value approximately
1 eV, $m_i \ltap 1$ eV.
The three neutrino mixing scheme is described
(to a good approximation)
by the Pontecorvo, Maki, Nakagawa, Sakata (PMNS) $3\times 3$
unitary mixing matrix, $U_{\rm PMNS}$. In the widely used
standard parametrization \cite{PDG12},  $U_{\rm PMNS}$ is
expressed in terms of the solar, atmospheric and reactor
neutrino mixing angles $\theta_{12}$,  $\theta_{23}$ and
$\theta_{13}$, respectively, and one Dirac  ($\delta$), and
two Majorana \cite{BHP80,Doi81}  ($\alpha$ and $\beta$)
CP violation (CPV) phases.
In this parametrization, the elements of the first row of the
PMNS matrix, $U_{ei}$, $i=1,2,3$, which play important role
in our further discussion, are given by
\beq
U_{e1} = c_{12}\,c_{13}\,,
~U_{e2} = s_{12}\,c_{13}\,e^{i\alpha}\,,
~U_{e3} = s_{13}\,e^{i(\beta - \delta)}\,,
\label{Uei}
\eeq
%
where we have used the standard notation
$c_{ij} = \cos\theta_{ij}$,
$s_{ij} = \sin\theta_{ij}$ with
$0\leq \theta_{ij}\leq \pi/2$,
$0\leq \delta \leq 2\pi$ and,
in the case of interest for our analysis
\footnote{Note that the
two Majorana phases $\alpha_{21}$
and  $\alpha_{31}$ defined in \cite{PDG12}
are twice the phases $\alpha$ and $\beta$:
$\alpha_{21} =2\alpha$, $\alpha_{31}=2\beta$.
},
 $0\leq \alpha,\beta \leq \pi$,
(see, however, \cite{EMSPEJP09}).
If CP invariance holds, we have
$\delta =0,\pi$, and \cite{LW81}
$\alpha,\beta = 0,\pi/2,\pi$.
%
%
%

 The neutrino oscillation data, accumulated over many years,
allowed to determine the parameters which drive the solar and
atmospheric neutrino oscillations,
$\Delta m^{2}_{\odot}\equiv \Delta m^{2}_{21}$, $\theta_{12}$ and
$|\Delta m_A^2| \equiv |\Delta m^{2}_{31}| \cong
|\Delta m^{2}_{32}|$, $\theta_{23}$,
with a high precision (see, e.g., \cite{Nu2012}).
Furthermore, there were  spectacular developments
in the last year in what concerns the angle $\theta_{13}$
(see, e.g., \cite{PDG12}). They culminated
in a high precision determination of $\sin^22\theta_{13}$
in the Daya Bay experiment using the reactor
$\bar{\nu}_e$ \cite{An:2012eh}:
\begin{equation}
 \sin^22\theta_{13} = 0.089 \pm 0.010 \pm 0.005\,.
\label{DBayth13}
\end{equation}
%
Similarly, the RENO,  Double Chooz, and T2K
experiments reported, respectively, $4.9\sigma$, $3.1\sigma$ and
$3.2\sigma$ evidences for a non-zero value of $\theta_{13}$
\cite{RENODCT2Kth13}, compatible with the Daya Bay result.

  A global analysis of the latest
neutrino oscillation data presented at
the Neutrino 2012 International Conference \cite{Nu2012}
was performed in \cite{Fogli:2012XY}.
We give below the best fit values of $\Delta m^2_{21}$,
$\sin^2\theta_{12}$, $|\Delta m^2_{31(32)}|$ and
$\sin^2\theta_{13}$, obtained in \cite{Fogli:2012XY},
which will be relevant for our further discussion:
\begin{eqnarray}
\label{Delta2131}
\Delta m^2_{21} = 7.54 \times 10^{-5} \ {\rm eV^2}\,,
& |\Delta m^2_{31(32)}| = 2.47~(2.46) \times 10^{-3} \ {\rm eV^2}\,,\\
\label{sinsq1213}
\sin^2\theta_{12} = 0.307,
& \sin^2\theta_{13} = 0.0241~(0.0244)\,,
\end{eqnarray}
%
where the values (the values in brackets) correspond to
NO (IO) neutrino mass spectrum. We will neglect the
small differences between the NO and IO values of
$|\Delta m^2_{31(32)}|$ and  $\sin^2\theta_{13}$
and will use $|\Delta m^2_{31(32)}| = 2.47\times 10^{-3} \ {\rm eV^2}$,
$\sin^2\theta_{13} = 0.024$ in our numerical
analysis.

 After the successful measurement of $\theta_{13}$,
the determination of  the absolute neutrino mass scale,
of the type of the neutrino mass spectrum,
of the nature of massive neutrinos,
as well as getting information about
the status of CP violation in the lepton sector,
remain the highest priority
goals of the research in neutrino physics.
Establishing whether CP is conserved or
not in the lepton sector
is of fundamental importance,
in particular, for making
progress in the understanding of
the origin of the matter-antimatter
asymmetry of the Universe
(see, e.g., \cite{fy-86,PPRio06,Branco2012}).

   Some time ago one of the present
authors proposed to use
atoms or molecules for systematic experimental
determination of the neutrino mass matrix
\cite{my-rnpe1,my-rnpe2}.
Atoms have a definite advantage
over conventional target of nuclei: their available
energies are much closer to neutrino masses.
The process proposed is cooperative de-excitation
of atoms in a metastable state.
For the single atom the process is
$|e\rangle \rightarrow |g\rangle + \gamma +(\nu_i + \nu_j)$,
$i,j=1,2,3$, where $\nu_i$'s are neutrino mass eigenstates.
If $\nu_i$ are Dirac fermions, $(\nu_i + \nu_j)$ should be
understood for $i=j$ as $(\nu_i + \bar{\nu}_i)$, and
as either  $(\nu_i + \bar{\nu}_j)$ or  $(\nu_j + \bar{\nu}_i)$
when $i\neq j$, $\bar{\nu}_i$ being the antineutrino with mass $m_i$.
If $\nu_i$ are Majorana particles,  we have $\bar{\nu}_i \equiv \nu_i$
and $(\nu_i + \nu_j)$ are the Majorana neutrinos with masses
$m_i$ and $m_j$.

\vspace{0.2cm}
  The proposed experimental method is to measure,
under irradiation of
two counter-propagating trigger lasers,
the continuous photon ($\gamma$)
energy spectrum below each of the six thresholds
$\omega_{ij}$ corresponding to the production of the six different
pairs of neutrinos, $\nu_1\nu_1$, $\nu_1\nu_2$,..., $\nu_3\nu_3$:
$\omega < \omega_{ij}$,
$\omega$ being  the photon energy, and \cite{my-rnpe1,my-rnpe2}
\begin{eqnarray}
&&
\omega_{ij} = \omega_{ji}
= \frac{\epsilon_{eg}}{2} - \frac{(m_i+m_j)^2}{2\epsilon_{eg}}\,,~~~
i,j=1,2,3,~~m_{1,2,3}\geq 0\,,
\label{thresholds}
\end{eqnarray}
%
where $\epsilon_{eg}$ is the energy difference between
the two relevant atomic levels.

\begin{figure*}[htbp]
 \begin{center}
 \epsfxsize=0.4\textwidth
 \centerline{\epsfbox{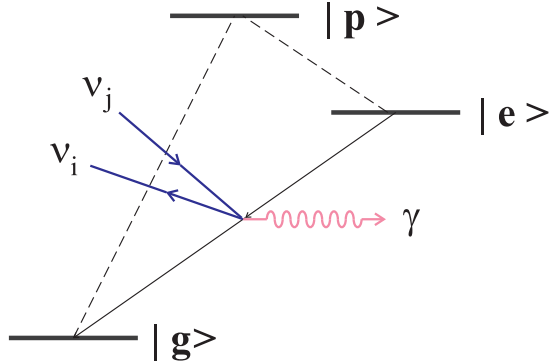}}
  \hspace*{\fill}
   \caption{$\Lambda-$type  atomic level for RENP
$|e \rangle \rightarrow |g\rangle + \gamma + \nu_i\nu_j$
with $\nu_i$ a neutrino mass eigenstate.
   Dipole forbidden transition
$|e \rangle \rightarrow |g\rangle + \gamma + \gamma$
may also occur via weak E1$\times$ M1 couplings to $|p\rangle$.
}
   \label{lambda-type atom for renp}
 \end{center}
\end{figure*}

The process occurs in the  3rd
order (counting the four Fermi weak interaction as the 2nd order) of electroweak theory
as a combined weak and QED process,
as depicted in Fig. \ref{lambda-type atom for renp}.
Its effective amplitude has the form of
\begin{eqnarray}
\label{renp amplitude}
&&
\langle g| \vec{d}|p \rangle\cdot\vec{E}
\frac{G_F\sum_{ij}a_{ij}\nu_j^{\dagger}\vec{\sigma}\nu_i}{\epsilon_{pg} -\omega}
\cdot\langle p | \vec{S}_e|e \rangle
\,,\\[0.25cm]
&&a_{ij} = U_{ei}^*U_{ej} - \frac{1}{2} \delta_{ij}\,,
\label{aij}
\end{eqnarray}
%
where $U_{ei}$, $i=1,2,3$, are the elements of
the first row of the neutrino mixing matrix $U_{PMNS}$,
given in eq. (\ref{Uei}).
The atomic part of the probability amplitude
involves three states $|e\rangle, |g\rangle , |p\rangle$,
where the two states $|e\rangle , |p\rangle$,
responsible for the neutrino pair emission,
are connected by a magnetic dipole type operator,
the electron spin $\vec{S}_e$. The
$|g\rangle - |p\rangle$ transition involves a
stronger electric dipole operator $\vec{d}$.
From the point of selecting candidate atoms,
E1$\times$M1 type transition must be chosen
between the initial and the final states ($|e\rangle$ and $ |g\rangle$).
The field $\vec{E}$ in eq. (\ref{renp amplitude}) is the one stored in
the target by the counter-propagating fields.
The formula has some similarity to the case of stimulated emission.
By utilizing the accuracy of trigger laser one can decompose,
in principle, all six photon energy thresholds at $\omega_{ij}$,
thereby resolving the neutrino mass eigenstates instead of the
flavor eigenstates.
The spectrum rise below each threshold  $\omega \leq \omega_{ij}$
depends, in particular, on  $|a_{ij}|^2$ and
is sensitive to the type of the neutrino mass spectrum,
to the nature of massive neutrinos, and, in the case
of  emission of two different Majorana neutrinos,
to the Majorana CPV phases in the neutrino mixing matrix
(see further).
%

  The disadvantage of atomic targets is their smallness of rates which
are very sensitive to available energy of order eV.
This can be overcome by developing, with the aid of a trigger laser,
macro-coherence of atomic polarization
to which the relevant amplitude is proportional,
as discussed in \cite{macro-coherence,psr dynamics}.
The macroscopic polarization supported by trigger field
gives rise to enhanced rate $\propto n^2 V$,
where $n$ is the number density of excited atoms and $V$
is the volume irradiated by the trigger laser.
The proposed atomic process may be called radiative emission of
neutrino pair, or RENP in short.
The estimated rate roughly of order mHz or a little less
makes it feasible to plan realistic RENP experiments
for a target number of order of the Avogadro number,
within a small region of order $1\sim 10^2$ cm$^3$,
if the rate enhancement works as expected.

 The new atomic process of RENP has a rich variety of
neutrino phenomenology, since there are six
independent thresholds for each
target choice, having a strength
proportional to different combinations of
neutrino masses and mixing parameters.
In the present work we shall correct
the spectrum formula for the Majorana neutrino
case given in \cite{my-rnpe2}
and also extend the discussion of the atomic spin factor.

 In the numerical
results presented here we
show the sensitivity of the RENP related
photon spectral shape
to various observables;
the absolute neutrino mass scale, the
type of neutrino mass spectrum,
the nature of massive of neutrinos and
the Majorana CPV phases in the case of
massive Majorana neutrinos.
All these observables can be determined
in one experiment, each observable with a different
degree of difficulty, once the RENP process is
experimentally established.
For atomic energy available in the
RENP process of the order of a fraction of eV,
the observables of interest can be ranked
in the order of increasing difficulty of
their determination as follows:\\
(1) The absolute neutrino mass scale, which can be
fixed by, e.g.,  measuring the smallest photon energy threshold
${\rm min}(\omega_{ij})$ near which the RENP rate is maximal:
${\rm min}(\omega_{ij})$ corresponds to the
production of a pair of the heaviest neutrinos
(${\rm max}(m_j) \gtap 50$ meV).\\
(2) The neutrino mass hierarchy, i.e., distinguishing
between the normal hierarchical (NH), inverted hierarchical (IH)
and quasi-degenerate (QD) spectra, or a spectrum with
partial hierarchy (see, e.g., \cite{PDG12}).\\
(3) The nature (Dirac or Majorana) of massive neutrinos.\\
(4) The measurement on the Majorana CPV phases if the massive neutrinos
are Majorana particles.

The last item is particularly challenging.
The importance of getting information about the Majorana CPV violation
phases in the proposed RENP experiment stems, in particular, from the
possibility that these phases play a fundamental
role in the generation of the baryon
asymmetry of the Universe \cite{PPRio06}.
The only other experiments which, in principle,
might provide information about the Majorana CPV phases are the
neutrinoless double beta ($\betabeta$-) decay
experiments (see, e.g., \cite{PPSchw0505,bb0nuExp2}).

  The paper is organized as follows.
In Section 2 the basic RENP spectral rate
formula is given along with
comments on how the Majorana vs Dirac distinction arises.
We specialize to rates under no magnetic field so that
the experimental setup is simplest.
In Section 3 we discuss the physics potential of a
RENP experiment for measuring the absolute neutrino
mass scale and determining the type of neutrino mass spectrum
(or hierarchy) and the nature (Dirac or Majorana)
of massive neutrinos. This is done on the examples of
a candidate transition of Yb
$J= 0$ metastable state and of a hypothetical atom
of scaled down energy of the transition in which
the photon and the two neutrinos are emitted.
Section 4 contains Conclusions.

\vspace{-0.5cm}
\section{Photon Energy Spectrum in RENP}

\vspace{-0.2cm}
When the target becomes macro-coherent by
irradiation of trigger laser,
RENP process conserves both the momentum and the energy
which are shared by a photon and two emitted neutrinos
resulting in the threshold relation (\ref{thresholds})
\cite{macro-coherence}.
The atomic recoil can be neglected to a good approximation.
Since neutrinos are practically impossible to measure,
one sums over neutrino momenta and helicities,
and derives the single photon  spectrum
as a function of photon energy $\omega$.
We think of experiments that do not apply magnetic
field and neglect effects of atomic spin orientation.
The neutrino helicity (denoted by $h_{r}\,, r=1,2$)
summation in the squared neutrino
current $j^{k}= a_{ij}\nu^{\dagger}_i \sigma_{k} \nu_j$
gives  bilinear terms of neutrino momenta
(see \cite{my-rnpe1} and the discussion
after eq. (\ref{Iomegamassless})):
\begin{eqnarray}
&&
K^S_{kn} \equiv
\sum_{h_1,h_2}  j^{k} (j^{n})^{\dagger}
\nonumber
\\ &&
= |a_{ij}|^2\left [
\left (1 - \delta_M\,\frac{m_i m_j}{E_i E_j}
\left (1 - 2\,\frac{({\rm Im}(a_{ij}))^2}{|a_{ij}|^2}\right )
 \right )\delta_{kn}
+ \frac{1}{E_i E_j}
\left (p_i^k p_j^n + p_j^k p_i^n - \delta_{kn}\vec{p}_i\cdot\vec{p}_j
\right )\right]
\,.
\label{helicity sum of current22}
\end{eqnarray}
%
The case $\delta_M = 1$ applies to Majorana neutrinos,
$\delta_M = 0$ corresponds to Dirac neutrinos.
The term $\propto m_im_j(1 - 2({\rm Im}(a_{ij}))^2/|a_{ij}|^2)$
is similar to, and has the same physical origin as,
the term $\propto M_iM_j$ in the production cross section
of two different Majorana neutralinos
$\chi_i$ and $\chi_j$
with masses $M_i$ and $M_j$ in the
process of $e^- + e^+ \rightarrow \chi_i + \chi_j$
\cite{eechi1chi286}. The term $\propto M_iM_j$
of interest determines, in particular, the threshold behavior
of the indicated cross section.

The subsequent neutrino momentum integration
(with $E_i = \sqrt{\vec{p}_i^2 + m_i^2}$ being the neutrino energy)
\begin{eqnarray}
&&
\int \frac{d^3p_1 d^3p_2}{(2\pi)^2} \delta^3
(\vec{k} + \vec{p}_1 + \vec{p}_2) \delta (\epsilon_{eg} - \omega - E_1 - E_2)
K^S_{ij} \equiv
\frac{1}{2\pi}\int d{\cal P}_{\nu}K^S_{ij}
\,,
\end{eqnarray}
%
can be written as a second rank tensor of photon
momentum,
\(\:
k_i k_jG^{(1)} + \delta_{ij}\vec{k}^2 G^{(2)}
\:\)
from rotational covariance.
Two coefficient functions $G^{(i)}$ are readily evaluated
by taking the trace $\sum_{i=j}$ and a product with $k_i k_j$
and using the energy-momentum conservation.
But their explicit forms are not necessary in subsequent computation.

We now consider sum over magnetic quantum numbers
of E1$\times$M1 amplitude squared:
\begin{eqnarray}
&&
R = \int d{\cal P}_{\nu}
\frac{\sum_{ M_e}}{2J_e+1}
\sum_{M_g}|\sum_{ M_p}
\langle gM_g|\vec{d}\cdot\vec{E} |pM_p\rangle \cdot\langle pM_p| \vec{S}_e
\cdot\vec{j}_{\nu} |eM_e\rangle|^2
\,.
\end{eqnarray}
%
The field $\vec{E}$ is assumed to be oriented
along the trigger axis taken  parallel  to $3-$axis.
Since there is no correlation of neutrino pair emission to
the trigger axis,
one may use the isotropy of space and replace
$(\vec{S}_e \cdot \vec{k})_{ni}(\vec{S}_e \cdot \vec{k})_{in'}$
by $(\vec{S}_e )_{ni}\cdot(\vec{S}_e )_{in'}\vec{k}^2/3$.
Using the isotropy,
we define the atomic spin factor $C_{ep}$(X) of X atom by
\begin{eqnarray}
&&
\frac{\sum_{ M_e}}{2J_e+1}
\langle pM_p| \vec{S}_e|eM_e\rangle\cdot
\langle eM_e| \vec{S}_e|pM_p'\rangle = \delta_{M_pM_p'}C_{ep}({\rm X})
\,.
\end{eqnarray}
%
This means that only the trace part of
eq. (\ref{helicity sum of current22}), $4 K^S_{ii}/3$,
is relevant for the neutrino phase space integration.

  The result is summarized by separating the
interference term relevant to the case of
Majorana neutrinos $\nu_i$:
\begin{eqnarray}
&&
\Gamma_{\gamma 2\nu}(\omega)  = \Gamma_0 I(\omega) \eta_{\omega}(t)
\,, \hspace{0.5cm}
\Gamma_0 =
\frac{ 3 n^2 V  G_F^2 \gamma_{pg} \epsilon_{eg}n}
{2\epsilon_{pg}^3 }(2J_p+1)C_{ep}
\,,
\label{rnpe spectrum rate gamma}
\\ &&
I(\omega) = \frac{1}{(\epsilon_{pg}-\omega)^2}
\sum_{ij}\,|a_{ij}|^2\,\Delta_{ij}(\omega)
\left ( I_{ij}(\omega) -
\delta_M m_i\, m_j\,B^M_{ij}\,
\right )
\,,
\label{rnpe spectrum rate00}
\\ &&
B^M_{ij} = \frac{\Re (a_{ij}^2)}{|a_{ij}|^2} =
\left (1 - 2\,\frac{({\rm Im}(a_{ij}))^2}{|a_{ij}|^2}\right )
\,, \hspace{0.3cm}
a_{ij} = U_{ei}^*U_{ej} - \frac{1}{2} \delta_{ij}
\,,
\label{rnpe spectrum rate01}
\\ &&
\Delta_{ij}(\omega)
=\frac{1}{\epsilon_{eg} (\epsilon_{eg} -2\omega) }\left\{
\left( \epsilon_{eg} (\epsilon_{eg} -2\omega) - (m_i + m_j)^2\right)
\left( \epsilon_{eg} (\epsilon_{eg} -2\omega) - (m_i - m_j)^2\right)
\right\}^{1/2}
\,,
\label{rnpe spectrum rate 3}
\\ &&
I_{ij}(\omega) =\left(
\frac{1}{3}\epsilon_{eg}(\epsilon_{eg}-2\omega)
+ \frac{1}{6}\omega^2
- \frac{1}{18}\omega^2 \Delta_{ij}^2(\omega)
- \frac{1}{6}(m_i^2+ m_j^2)
- \frac{1}{6}\frac{(\epsilon_{eg}-\omega)^2}
{\epsilon_{eg}^2(\epsilon_{eg}-2\omega)^2}(m_i^2 - m_j^2)^2
\right)
\,.
\nonumber  \\ &&
\label{rnpe spectrum rate 1}
\end{eqnarray}
%
The term $\propto \delta_{M}\,m_im_j$ appears only for the Majorana case. We shall define and
discuss the dynamical dimensionless factor $\eta_{\omega}(t)$ further below. The limit of massless
neutrinos gives the spectral form,
\begin{eqnarray}
&&
I(\omega; m_i = 0) =
\frac{\omega^2 - 6\epsilon_{eg}\omega +
3\epsilon_{eg}^2}{12(\epsilon_{pg}-\omega)^2}
\,,
\label{Iomegamassless}
\end{eqnarray}
%
where the prefactor of $\sum_{ij}|a_{ij}|^2 =3/4$ is
calculated using the unitarity of the neutrino mixing matrix.
On the other hand, near the threshold
these functions have the behavior
$\propto \sqrt{\omega_{ij}-\omega}$.

We will explain next the origin of the interference
term for Majorana neutrinos. The two-component Majorana neutrino
field can be decomposed in terms of plane wave modes as
\begin{eqnarray}
&&
\psi^M(\vec{x},t) =
\sum_{i, \vec{p}}
\left( u( \vec{p}) e^{-iE_i t + i\vec{p}\cdot\vec{x}} b_i( \vec{p})
+ u^c( \vec{p})e^{iE_i t - i\vec{p}\cdot\vec{x}}   b_i^{\dagger}( \vec{p})
\right)
\,,
\end{eqnarray}
%
where the annihilation $b_i( \vec{p})$ and creation $b^{\dagger}_i( \vec{p})$
operators appears as a conjugate pair of the same type of operator $b$
 in the expansion
(the index $i$ gives the $i-$th neutrino of mass $m_i$, and the helicity
summation is suppressed for simplicity).
The concrete form of the 2-component conjugate wave
function $u^c \propto i\sigma_2 u^*$
is given in \cite{my-rnpe1}.
A similar expansion can be written in terms of four component
field if one takes into account the chiral projection
$(1-\gamma_5)/2$ in the interaction.
The Dirac case is different involving different type of operators
$b_i(\vec{p})$ and $d^{\dagger}_i(\vec{p})$:
\begin{eqnarray}
&&
\psi^D(\vec{x},t) = \sum_{i, \vec{p}} \left( u(\vec{p}) e^{-iE_i t + i\vec{p}\cdot\vec{x}} b_i(\vec{p})
+ v(\vec{p}) e^{iE_i t - i\vec{p}\cdot\vec{x}}  d_i^{\dagger}(\vec{p})
\right)
\,.
\end{eqnarray}
Neutrino pair  emission amplitude of modes $i\vec{p}_1, j\vec{p}_2$ contains two terms
in the case of Majorana particle:
\begin{eqnarray}
&&
b_i^{\dagger}b_j^{\dagger}
\left( a_{ij} u^*(\vec{p}_1)u^c(\vec{p}_2) - a_{ji}u^*(\vec{p}_2)u^c(\vec{p}_1) \right)
\,,
\end{eqnarray}
and its rate involves
\begin{eqnarray}
&&
\hspace*{-1cm}
 \frac{1}{2}\left| a_{ij} u^*(\vec{p}_1)u^c(\vec{p}_2) - a_{ji}u^*(\vec{p}_2)u^c(\vec{p}_1) \right|^2
=  \frac{1}{2}|a_{ij}|^2\left( |\psi(1,2)|^2 + |\psi(2,1)|^2 \right)
- \Re (a_{ij}^2)\left( \psi(1,2) \psi(2,1)^*
\right)
\,, \nonumber \\ &&
\end{eqnarray}
where the relation $a_{ji} = a_{ij}^*$ is used and $\psi(1,2)  = u^*(\vec{p}_1)u^c(\vec{p}_2)$.
The result of the helicity sum \\
$\sum \left( \psi(1,2) \psi(2,1)^*\right)$
is in \cite{my-rnpe1}, which then
gives the interference term $\propto B_{ij}^M$ in
the formula (\ref{rnpe spectrum rate01}).

\vspace{0.3cm}
  We see from eqs. (\ref{rnpe spectrum rate gamma}) and
(\ref{rnpe spectrum rate00}) that
the overall decay rate is determined by the
energy independent $\Gamma_0$, while
the spectral information is in the dimensionless function $I(\omega)$.
The rate $\Gamma_0$ given here is obtained by replacing the field
amplitude $E$ of Eq.\ref{renp amplitude} squared
by $\epsilon_{eg}n$, which is the atomic energy density
stored in the upper level $|e\rangle$.

The dynamical factor $\eta_{\omega}(t)$ is defined by
a space integral of a product of macroscopic polarization
squared times field strength, both in dimensionless units,
\begin{eqnarray}
&&
\eta_{\omega}(t) = \frac{1}{\alpha_m L}\int_{-\alpha_m L/2}^{\alpha_m L/2}d\xi
\frac{r_1(\xi\,, \alpha_m t)^2 + r_2(\xi\,, \alpha_m t)^2 }{4}|e(\xi\,, \alpha_m t)|^2
\,.
\label{spatial integral}
\end{eqnarray}
Here $r_1 \pm i r_2$ is the medium polarization normalized to
the target number density.

The dimensionless field strength $|e(\xi,\tau)|^2 =
|E(\xi=\alpha_m x,\tau=\alpha_m t)|^2/(\epsilon_{eg}n)$
is to be calculated using the evolution equation for
field plus medium polarization in \cite{psr dynamics},
where $\xi = \alpha_m x$ ($\alpha_m = \epsilon_{eg}\mu_{ge}n/2$
with $\mu_{ge}$ the off-diagonal coefficient of AC Stark
shifts \cite{my-rnpe2}) is the atomic site position in
dimensionless unit  along the trigger laser
direction ($-L/2 <x < L/2$ with $L$ the target length), and
$\tau = \alpha_m t$ is the dimensionless time.
The characteristic unit
of length and time are
$\alpha_m^{-1} \sim (1 {\rm cm}) (n/10^{21}$cm$^{-3})^{-1}$
and ($40 $ps$)(n/10^{21}$cm$^{-3})^{-1}$ for Yb discussed below.
We expect that $\eta_{\omega}(t)$ in the formula given
above is roughly of order unity or less.\footnote{There is a weak dependence of the dynamical factor $\eta_{\omega}(t)$
on the photon energy $\omega$, since the field $e$
in Eq.\ref{spatial integral}, a solution of the evolution equation,
is obtained for the initial boundary condition of
frequency $\omega$ dependent trigger laser irradiation.
} We shall have more comments on this at the end of this section.

  Note that what we calculate here is not
the differential spectrum at
each frequency, instead it is the spectral
rate of number of events per unit time at each photon energy.
Experiments for the same target atom are
repeated at different frequencies
$\omega_1 \leq \omega_{11}$ in the NO case
(or $\omega_1 \leq \omega_{33}$ in the IO case) since
it is irradiated by
two trigger lasers of different frequencies of
$\omega_i$ (constrained by $\omega_1 + \omega_2 =
\epsilon_{eg}$) from counter-propagating directions.

As a standard reference target we take Yb atom
and the following de-excitation path,
\begin{eqnarray}
&&
{\rm Yb}; \hspace{0.3cm}
|e \rangle = (6s6p)\,^3P_0
\,, \hspace{0.3cm}
|g \rangle = (6s^2)\, ^1S_0
\,, \hspace{0.3cm}
|p \rangle = (6s6p)\,^3P_1
\,.
\end{eqnarray}
%
The relevant atomic parameters are as follows \cite{nist}:
\begin{eqnarray}
&&
\epsilon_{eg} =  2.14349~{\rm eV}
\,, \hspace{0.3cm}
\epsilon_{pg} = 2.23072~{\rm eV}
\,, \hspace{0.3cm}
\gamma_{pg} = 1.1~{\rm MHz}
\,.
\end{eqnarray}
%
The notation based on $LS$ coupling is used
for Yb electronic configuration, but this approximation
must be treated with care, since there might be
a sizable mixing based on $jj$ coupling scheme.
The relevant atomic spin factor $C_{ep}$(Yb) is estimated, using
the spin Casimir operator within an irreducible
representation of $LS$ coupling. Namely,
\begin{eqnarray}
&&
\langle ^3P_0|\vec{S}_e|^3P_1, M\rangle \cdot \langle ^3P_1, M|\vec{S}_e|^3P_0 \rangle
= \frac{1}{3}\sum_M \langle ^3P_0|\vec{S}_e|^3P_1, M\rangle \cdot \langle ^3P_1, M|\vec{S}_e|^3P_0 \rangle
= \frac{2}{3}
\,,
\end{eqnarray}
%
since $\vec{S}_e \cdot\vec{S}_e = 2$
for the spin triplet.
This gives $C_{ep}$(Yb)$= 2/3$ for the intermediate path chosen.

We also considered another path, taking the intermediate state of Yb,
$ ^1P_1$ with $\epsilon_{pg} = 3.10806~{\rm eV} \,,$ \\
$\gamma_{pg} = 0.176~{\rm GHz}$.
Using a theoretical estimate of A-coefficient $4.6 \times 10^{-2}$ Hz
for $^1P_1 \rightarrow ^3P_1$ transition given in NIST \cite{nist}
and taking the estimated Lande g-factor \cite{atomic physics},
3/2 for the $^3P_1$ case,
we calculate the mixed fraction of $jj$ coupling
scheme in $LS$ forbidden amplitude squared
$|\langle ^1P_1|\vec{S}_e|^3P_1\rangle|^2$,
to give  $C_{ep} \sim 1 \times 10^{-4}$.

  Summarizing, the overall rate factor $\Gamma_0$ is given by
\begin{eqnarray}
&&
\Gamma_0 = \frac{ 3 n^2 V  G_F^2 \gamma_{pg} \epsilon_{eg}n}{2\epsilon_{pg}^3 }
(2J_p + 1) C_{ep}
\sim 0.37~{\rm mHz} (\frac{n}{10^{21}~{\rm cm}^{-3}})^3
\frac{V}{10^2~{\rm cm}^3}\,,
\end{eqnarray}
%
where the number is valid for the Yb first excited state of $J=0$.
If one chooses the other intermediate
path, $^1P_1$, the rate $\Gamma_0$ is estimated to be of order,
$1\times 10^{-3}$ mHz,
a value much smaller than that of the $^3P_1$ path.
The denominator factor $1/(\epsilon_{pg}-\omega)^2$ is slightly
larger for the $^3P_1$ path, too.
We consider the intermediate $^3P_1$ path alone in the following.

The high degree of sensitivity to the target number density $n$
seems to suggest that solid environment is the best choice.
But de-coherence in solids is fast, usually sub-picoseconds, and one has
to verify how efficient coherence development is achieved in
the chosen target.

Finally,
we discuss a stationary value of time independent
$\eta_{\omega}(t)$ (\ref{spatial integral})
some time after trigger irradiation.
The stationary value may arise when many soliton pairs of
absorber-emitter \cite{psr dynamics} are
created, since the target in this stage is expected  not to emit photons
of PSR origin (due to the macro-coherent $|e\rangle \rightarrow |g\rangle + \gamma \gamma$),
or emits very little only at target ends, picking up
an exponentially small leakage tail.
This is due to the stability of solitons against
two photon emission.
Thus the PSR background is essentially negligible.
According to \cite{soliton dynamics},
the $\eta_{\omega}(t)$ integral (\ref{spatial integral})
is time dependent in general.
Its stationary standard reference value may be obtained
by taking the field from a single created soliton.
This quantity depends on target parameters such as $\alpha_m$ and
relaxation times.
Moreover, a complication arises, since many solitons may be created within the target,
and the number of created solitons should be multiplied in the rate.
This is a dynamical question that has to be addressed separately.
In the following sections we compute  spectral rates,
assuming $\eta_{\omega}(t) =1$.

\section{Sensitivity of the Spectral Rate to
Neutrino Mass Observables
and the Nature of Massive Neutrinos}

  We will discuss in what follows the potential of an
RENP experiment to get information about the absolute neutrino
mass scale, the type of the neutrino mass spectrum
and the nature of massive neutrinos.
We begin by recalling that
the existing data do not allow one to
determine the sign of
$\Delta m^2_{\rm A} = \Delta m^2_{31(2)}$
and in the case of 3-neutrino mixing,
the two possible signs of
$\Delta m^2_{31(2)}$ corresponding to two
types of neutrino mass spectrum.
In the standard convention \cite{PDG12}
the two spectra read:\\
{\it i) spectrum with normal ordering (NO)}: $m_1 < m_2 < m_3$,
$\Delta m^2_{\rm A} = \Delta m^2_{31} >0$,
$\Delta m^2_{21} > 0$,
$m_{2(3)} = (m_1^2 + \Delta m^2_{21(31)})^{1\over{2}}$;~~
{\it ii) spectrum with inverted ordering (IO)}:
$m_3 < m_1 < m_2$, $\Delta m^2_{\rm A} = \Delta m^2_{32}< 0$,
$\Delta m^2_{21} > 0$,
$m_{2} = (m_3^2 + \Delta m^2_{23})^{1\over{2}}$,
$m_{1} = (m_3^2 + \Delta m^2_{23} - \Delta m^2_{21})^{1\over{2}}$.
Depending on the values of the smallest neutrino mass,
${\rm min}(m_j) \equiv m_0$, the neutrino mass spectrum can also be
normal hierarchical (NH), inverted hierarchical (IH)
and quasi-degenerate (QD):
\begin{eqnarray}
\label{NH}
{\rm NH}:~&& m_1 \ll m_2 < m_3\,,
~m_2 \cong (\Delta m^2_{21})^{1\over{2}}\cong 0.009 ~{\rm eV}\,,~
m_3 \cong (\Delta m^2_{31})^{1\over{2}} \cong 0.05 ~{\rm eV}\,,\\
\label{IH}
{\rm IH}:~&& m_3 \ll m_1 < m_2\,,~
m_{1,2} \cong |\Delta m^2_{32}|^{1\over{2}}\cong 0.05~{\rm eV}\,,\\
\label{QD}
{\rm QD}:~&& m_1 \cong m_2 \cong m_3 \cong m\,,~
m_j^2 \gg |\Delta m^2_{\rm 31(32)}|\,,~ m \gtap 0.10~{\rm eV}\,.
\end{eqnarray}
%
All three types of spectrum are compatible
with the existing constraints on the absolute scale
of neutrino masses $m_j$.
\begin{table}
\centering \caption{\label{tab_aij}
The quantity $|a_{ij}|^2=|U^*_{ei}U_{ej}-\frac{1}{2}\delta_{ij}|^2$}
   \begin{tabular}{|c|c|c|}
       \hline\noalign{\vskip0.2cm}
      $|a_{11}|^2=|c_{12}^2c_{13}^2-\frac{1}{2}|^2$
& $|a_{12}|^2=c_{12}^2s_{12}^2c_{13}^4$
& $|a_{13}|^2=c_{12}^2s_{13}^2c_{13}^2$
      \tabularnewline[0.2cm]
      \hline\noalign{\vskip0.2cm}
       $0.0311$                                      & $0.2027$                              & $0.0162$
       \tabularnewline[0.2cm]\hline
       \hline\noalign{\vskip0.2cm}
$|a_{22}|^2=|s_{12}^2c_{13}^2-\frac{1}{2}|^2$
& $|a_{23}|^2=s_{12}^2s_{13}^2c_{13}^2$
& $|a_{33}|^2=|s_{13}^2-\frac{1}{2}|^2$
      \tabularnewline[0.2cm]
      \hline\noalign{\vskip0.2cm}
      $0.0405$                                      & $0.0072$                              & $0.2266$
      \tabularnewline[0.2cm] \hline
   \end{tabular}
\end{table}
%
%
\subsection{General features of the Spectral Rate}
%

  The first thing to notice is that the
rate of emission of a given pair of neutrinos $(\nu_i + \nu_j)$ is
suppressed, in particular, by the factor $|a_{ij}|^2$,
independently of the nature of massive neutrinos.
The expressions for the six different factors
$|a_{ij}|^2$ in terms of the sines and cosines of
the mixing angles $\theta_{12}$ and $\theta_{13}$,
as well as their values corresponding to the best fit
values of $\sin^2\theta_{12}$ and
$\sin^2\theta_{13}$ quoted in eq. (\ref{Delta2131}),
are given in Table \ref{tab_aij}.
It follows from Table \ref{tab_aij} that
the least suppressed by the factor $|a_{ij}|^2$
is the emission of the pairs
$(\nu_3 +\nu_3)$ and $(\nu_1 +\nu_2)$, while the most
suppressed is the emission of $(\nu_2 +\nu_3)$.
The values of  $|a_{ij}|^2$ given in Table
\ref{tab_aij} suggest that in order to
be able to identify the emission of
each of the six pairs of neutrinos, the photon
spectrum, i.e., the RENP spectral rate, should be measured
with a relative precision not worse than
approximately $5\times 10^{-3}$.

  As it follows from eqs. (\ref{rnpe spectrum rate00}) and
(\ref{rnpe spectrum rate01}),
 the rate of emission of a pair of Majorana neutrinos with masses
$m_i$ and $m_j$ differs from the rate of emission of a pair of
Dirac neutrinos with the same masses by the interference term
$\propto m_im_j B^{M}_{ij}$. For $i=j$ we have $B^{M}_{ij} = 1$,
the interference term is negative and tends to suppress
the neutrino emission rate. In the case of $i \neq j$,
the factor $B^{M}_{ij}$, and thus the rate
of emission of a pair of different Majorana
neutrinos, depends on specific combinations
of the Majorana and Dirac CPV phases of the
neutrino mixing matrix: from
eqs. (\ref{rnpe spectrum rate01}) and
(\ref{Uei}) we get
\beq
 B^{M}_{12} = \cos 2\alpha\,,~~
B^{M}_{13} = \cos 2(\beta - \delta)\,,~~
B^{M}_{23} = \cos 2(\alpha - \beta + \delta)\,.
\label{BMij}
\eeq
%
In contrast, the rate of emission of a pair
of Dirac neutrinos  does not depend on
the CPV phases of the PMNS matrix.
In the case of CP invariance we have
$\alpha,\beta = 0,\pi/2,\pi$, $\delta = 0,\pi$, and,
correspondingly, $B^{M}_{ij} = -1~{\rm or}~+1$, $i\neq j$.
For  $B^{M}_{ij} = +1$, the interference term
tends to suppress
the neutrino emission rate, while for
$B^{M}_{ij} = -1$
it tends to increase it.
If some of the three relevant (combinations of)
CPV phases, say $\alpha$, has a CP violating value,
we would have $-1 <  B^{M}_{12} < 1$;
if all three are CP violating, the inequality
will be valid for each of the three factors
$B^{M}_{ij}$: $-1 <  B^{M}_{ij} < 1$, $i\neq j$.
Note, however, that the rates of emission of
$(\nu_1 + \nu_3)$ and of  $(\nu_2 + \nu_3)$
are suppressed by $|a_{13}|^2 = 0.016$ and
$|a_{23}|^2 = 0.007$, respectively. Thus,
studying the rate of emission of $(\nu_1 + \nu_2)$
seems the most favorable approach
to get information about the Majorana phase
$\alpha$,  provided the corresponding
interference term $\propto m_1m_2B^{M}_{12}$
is not suppressed by the smallness
of the factor $m_1m_2$. The mass $m_1$ can be very small
or even zero in the case of
NH neutrino mass spectrum, while for the
IH spectrum we have
 $m_1m_2\gtap |\Delta m^2_{32}|\cong 2.5\times 10^{-3}~{\rm eV^2}$ .
We note that all three of the
CPV phases in eq. (\ref{BMij})
enter into the expression for the
$\betabeta-$ decay effective  Majorana mass
as their linear combination
(see, e.g., \cite{PPSchw0505,BiPet87}):
\begin{eqnarray}
&&
|\sum_i m_i U_{ei}^2|^2 =
m_3^2 s_{13}^4 + m_2^2 s_{12}^4 c_{13}^4
 + m_1^2c_{12}^4 c_{13}^4
\nonumber \\ &&
\hspace*{-1cm}
+ 2m_1 m_2s_{12}^2 c_{12}^2 c_{13}^4 \cos (2\alpha)
 + 2m_1m_3 s_{13}^2 c_{12}^2 c_{13}^2\cos 2(\beta-\delta)
+ 2m_2m_3 s_{13}^2s_{12}^2 c_{13}^2
 \cos2(\alpha-\beta+\delta)
\,.
\end{eqnarray}
%

\begin{figure*}[htbp]
 \begin{center}
\epsfxsize=0.6\textwidth
\vskip -0.4cm
\centerline{\epsfbox{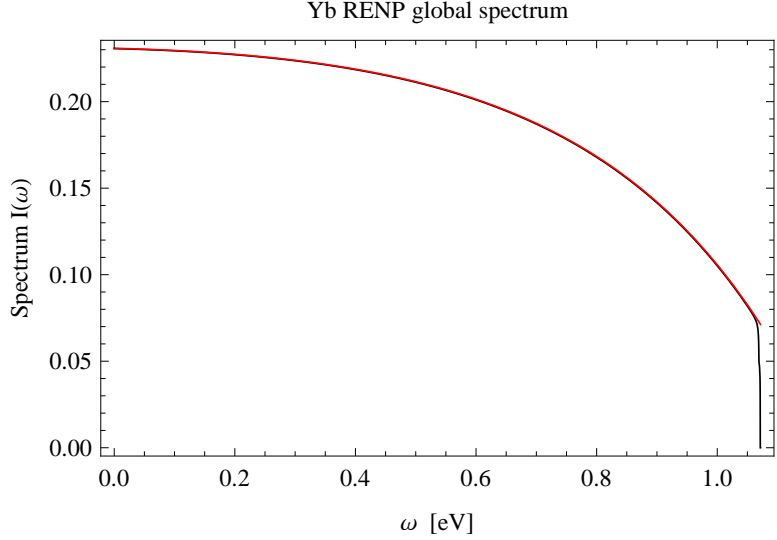}} \hspace*{\fill}
   \caption{Global feature of photon energy spectrum $I(\omega)$
for the  $^3P_0 \rightarrow {^1S_0}$ transitions in Yb.
The lines corresponding to $m_0 = 20$ meV (black lines)
and to massless neutrinos, $m_i = 0$ (red line),
are practically indistinguishable in this figure
(see text for details).
}
   \label{yb global}
 \end{center}
\end{figure*}
%
 In the case of $m_1 < m_2 < m_3$ (NO spectrum),
the ordering of the threshold energies at $\omega_{ij}=\omega_{ji}$
is the following: $\omega_{11} > \omega_{12} >
\omega_{22} > \omega_{13} > \omega_{23} > \omega_{33}$.
For NH spectrum with negligible $m_1$ which can be set to zero,
the factors $(m_i + m_j)^2 \equiv \kappa_{ij}$ in the
expression (\ref{thresholds}) for the threshold energy $\omega_{ij}$
are given by:
$\kappa_{11} = 0$,
$\kappa_{12} = \Delta m^2_{\rm 21}$,
$\kappa_{22} = 4 \Delta m^2_{\rm 21}$,
$\kappa_{13}= \Delta m^2_{\rm 31}$,
$\kappa_{23} = (\sqrt{\Delta m^2_{\rm 31}} + \sqrt{\Delta m^2_{\rm 21}})^2$,
$\kappa_{33} =  4 \Delta m^2_{\rm 31}$.
It follows from eq.(\ref{thresholds}) and the expressions for
$\kappa_{ij}$ that  $\omega_{11}$, $\omega_{12}$ and $\omega_{22}$ are
very close,  $\omega_{13}$ and $\omega_{23}$ are somewhat
more separated and the separation is the largest between
 $\omega_{22}$ and $\omega_{13}$, and  $\omega_{23}$ and $\omega_{33}$:
\begin{eqnarray}
\label{NH111222}
{\rm NH}:&&
\omega_{11} - \omega_{12} = \frac{1}{3}\,(\omega_{12} - \omega_{22}) =
\frac{1}{2\epsilon_{eg}}\,\Delta m^2_{21} \cong
1.759~(8.794)\times 10^{-5}~{\rm eV}\,,\\
\label{NH1323}
{\rm NH}:&&
\omega_{13} - \omega_{23} = \frac{1}{2\epsilon_{eg}}\,
(2\,\sqrt{\Delta m^2_{21}}\,\sqrt{\Delta m^2_{31}} +
\Delta m^2_{21})
\cong 0.219~(1.095)\times 10^{-3}~{\rm eV}\,,\\
\label{NH1323}
{\rm NH}:&&
\omega_{22} - \omega_{13} = \frac{1}{2\epsilon_{eg}}\,
(\Delta m^2_{31} - 4\Delta m^2_{21}) \cong
0.506~(2.529)\times 10^{-3}~{\rm eV}\,,~\\
\label{NH2333}
{\rm NH}:&&
\omega_{23} - \omega_{33} =
\frac{1}{2\epsilon_{eg}}
(3\Delta m^2_{31} - 2\,\sqrt{\Delta m^2_{21}}\,\sqrt{\Delta m^2_{31}} -
\Delta m^2_{21}) \cong 1.510~(7.548) \times 10^{-3}~{\rm eV}\,,
\end{eqnarray}
%
where the numerical values correspond to
$\Delta m^2_{21}$ given in eq. (\ref{Delta2131}) and
$\epsilon_{eg} =$ 2.14349~(
numbers in parenthesis corresponding
to the 1/5 of Yb value, namely 0.42870) eV.
We get similar results in what concerns the separation between
the different thresholds in the case of QD spectrum
and $\Delta m^2_{31} > 0$:
\begin{eqnarray}
\label{QD111222}
{\rm QD}:&&
\omega_{11} - \omega_{12} \cong \omega_{12} - \omega_{22} \cong
\omega_{13} -  \omega_{23}   \cong
\frac{1}{\epsilon_{eg}}\,\Delta m^2_{21} \cong
3.518~(17.588)\times 10^{-5}~{\rm eV}\,,\\
\label{QD1323}
{\rm QD}:&&
\omega_{22} - \omega_{13} \cong \omega_{23} - \omega_{33} -
\frac{1}{\epsilon_{eg}}\, \Delta m^2_{21} =
\frac{1}{\epsilon_{eg}}\,
(\Delta m^2_{31} - 2 \Delta m^2_{21})
\cong 1.082~(5.410)\times 10^{-3}~{\rm eV}\,.
\end{eqnarray}
%
 For spectrum with inverted ordering, $m_3 < m_1 < m_2$,
the ordering of the threshold energies
is different: $\omega_{33} > \omega_{13} >
\omega_{23} > \omega_{11} > \omega_{12} > \omega_{22}$.
In the case of IH spectrum with negligible $m_3 = 0$,
\begin{figure}
 \begin{center}
\vskip -0.4cm
\includegraphics[width=9.5cm,height=6.5cm]{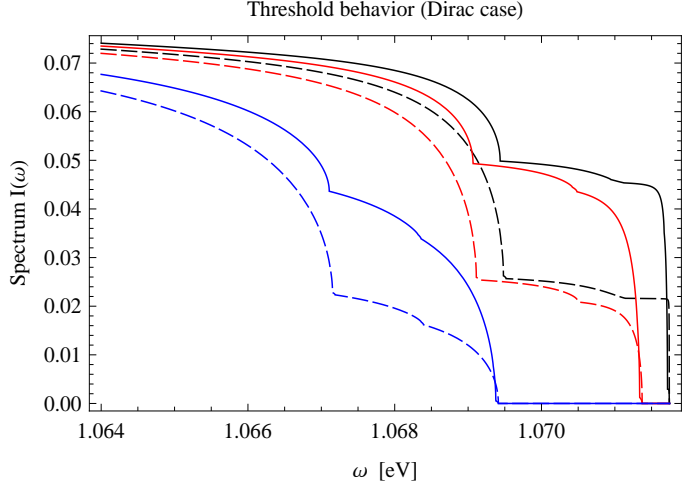}
   \caption{Photon energy spectrum from Yb
$^3P_0 \rightarrow {^1S_0}$ transitions in the threshold region
in the cases of NH spectrum (solid lines) and IH spectrum
(dashed lines) and for 3 different sets of Dirac neutrinos masses
corresponding to $m_0 = 2$ meV (black lines),
20 meV (red lines) and 50 meV (blue lines).
}
   \label{yb threshold}
 \end{center}
\end{figure}
%
we have:
$\kappa_{33} = 0$,
$\kappa_{13} =   \Delta m^2_{\rm 23} - \Delta m^2_{\rm 21}$,
$\kappa_{23} = \Delta m^2_{\rm 23}$,
$\kappa_{11}= 4 (\Delta m^2_{\rm 23} - \Delta m^2_{\rm 21})$,
$\kappa_{12} = (\sqrt{\Delta m^2_{\rm 23}} +
\sqrt{\Delta m^2_{\rm 23} - \Delta m^2_{\rm 21}})^2$,
$\kappa_{22} =  4\Delta m^2_{\rm 23}$.
Now not only $\omega_{11}$, $\omega_{12}$ and $\omega_{22}$, but also
$\omega_{13}$ and  $\omega_{23}$, are very close,
the corresponding  differences being all
$\sim \Delta m^2_{\rm 21}/\epsilon_{eg}$.
The separation between the thresholds $\omega_{33}$  and $\omega_{13}$,
and between $\omega_{23}$ and $\omega_{11}$, are considerably
larger, being $\sim  \Delta m^2_{\rm 23}/\epsilon_{eg}$.
These results remain valid also in the case of QD spectrum
and  $\Delta m^2_{32} < 0$.

 It follows from the preceding discussion that
in order to observe and determine all six
threshold energies $\omega_{ij}$, the photon energy $\omega$
should be measured with a precision not worse than
approximately $10^{-5}$ eV.
This precision is possible in our RENP experiments since
the energy resolution in the spectrum is determined by
accuracy of the trigger laser frequency, which is much
better than $10^{-5}$ eV.

%
\subsection{Neutrino Observables}
%

We will concentrate in what follows on the analysis of
the dimensionless spectral function $I(\omega)$ which
contains all the neutrino physics information of interest.

  In Fig. \ref{yb global} we show the global features
of the photon energy spectrum for the Yb $^3P_0 \rightarrow {^1S_0}$
transition in the case of massive Dirac neutrinos and
NH and IH spectra. For  $m_0 \leq $ 20 meV,
all spectra (including those corresponding to massive
Majorana neutrinos which are not plotted) look degenerate
owing to the horizontal and vertical axes scales used
to draw the figure.\\

{\bf The Absolute Neutrino Mass Scale.}
Much richer physics information is contained in
the spectrum near the thresholds $\omega_{ij}$.
Figure \ref{yb threshold} shows the Dirac neutrino
spectra for three different sets of values of the neutrino
masses  (corresponding to the smallest mass $m_{0}=2,\; 20,\; 50$ meV)
and for both the NO ($\Delta m^2_{31(32)} > 0$) and
IO ($\Delta m^2_{31(32)} < 0$) neutrino mass spectra.
One sees that the locations of the thresholds
corresponding to the three values of $m_0$
(and that can be seen in the figure)
differ substantially.
This feature can be used to
determine the absolute neutrino mass scale,
including the smallest mass, as evident in
differences of spectrum shapes for different masses of
$m_0$, 2,~20,~50 meV in Fig.\ref{yb threshold}.
In particular, the smallest mass can be determined by
locating the highest threshold ($\omega_{11} $ for NO and
$\omega_{33} $ for IO).
Also the location of the most prominent kink, which comes from
the heavier neutrino pair emission thresholds
($\omega_{33}$ in the NO case
and $\omega_{12}$
in the IO case), can
independently be used to extract the smallest neutrino mass value,
and thus to check consistency of two experimental methods.

If the spectrum is of the NO type, the measurement
of the position of the kink will determine the
value of $\omega_{33}$ and therefore of $m_3$.
 For the IO spectrum, the threshold  $\omega_{12}$ is very
close to the thresholds $\omega_{22}$  and $\omega_{11}$.
The rates of emission of the pairs $(\nu_2 + \nu_2)$
and $(\nu_1 + \nu_1)$, however, are smaller approximately
by the factors 10.0 and 12.7, respectively,
than the rate of emission of $(\nu_1 + \nu_2)$.
Thus, the kink due to the
$(\nu_1 + \nu_2)$ emission will be the easiest to observe.
The position of the kink will allow to determine
$(m_1 + m_2)^2$ and thus the absolute neutrino
mass scale. If the kink due to the emission of
$(\nu_2 + \nu_2)$ or  $(\nu_1 + \nu_1)$
will also be observed, it can be used
for the individual $m_1, m_2$
determination as well.\\

{\bf The Neutrino Mass Spectrum (or Hierarchy).}
Once the absolute neutrino mass scale is determined,
the distinction between the NH (NO) and IH (IO)
spectra can be made by measuring the ratio of
rates below and above the thresholds
$\omega_{33}$ and $\omega_{12}$ (or $\omega_{11}$),
respectively. We note that both of these
measurements can be done without knowing the
absolute counting rates.
For $m_0 \ltap 20$ meV and NH (IH) spectrum,
the ratio of the rates at $\omega$ just above the
$\omega_{33}$ ($\omega_{11}$) threshold and sufficiently
far below the indicated thresholds, $\tilde{R}$, is given by:
\begin{eqnarray}
\label{NHR}
{\rm NH}:&& \tilde{R}(\omega_{33};NH)
\cong \frac{\sum_{i,j} |a_{ij}|^2 - |a_{33}|^2}
    { \sum_{i,j} |a_{ij}|^2} \cong 0.70\,,\\
\label{IHR}
{\rm IH}:&& \tilde{R}(\omega_{11};IH) \cong
 \frac{|a_{33}|^2 + 2\,(|a_{13}|^2 +(|a_{23}|^2)}
     { \sum_{i,j} |a_{ij}|^2} \cong 0.36\,.
\end{eqnarray}
%
\begin{figure}
 \begin{center}
\includegraphics[width=9.5cm,height=6.5cm]{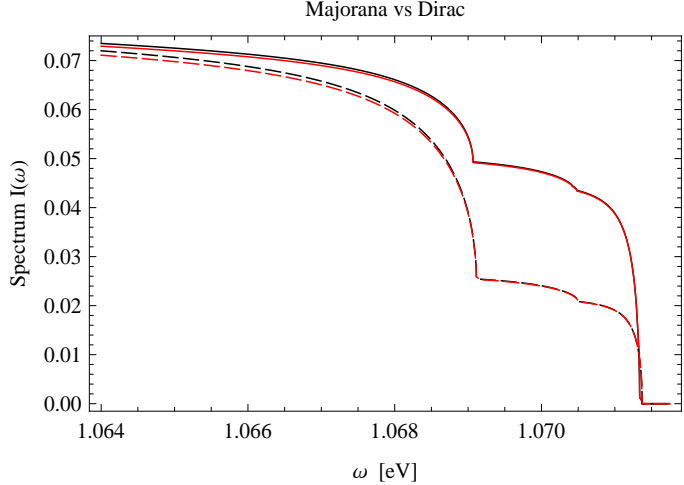}
  \caption{Spectra from Yb $^3P_0 \rightarrow {^1S_0}$
transitions in the cases of Dirac neutrinos
(black lines) and Majorana neutrinos (red lines)
with masses corresponding to $m_{0}=20$ meV, for
NH spectrum (solid lines) and IH spectrum (dashed lines).
}
   \label{Yb-Dirac-Majorana}
 \end{center}
\end{figure}
%
In obtaining the result (\ref{IHR}) in the IH case
we have assumed that $\omega_{22}$ and $\omega_{12}$
are not resolved, but the
kink due to the $\omega_{11}$
threshold could be observed.
\begin{figure}[t]
\begin{center}
\begin{tabular}{cc}
\includegraphics[width=7.5cm,height=6.5cm]{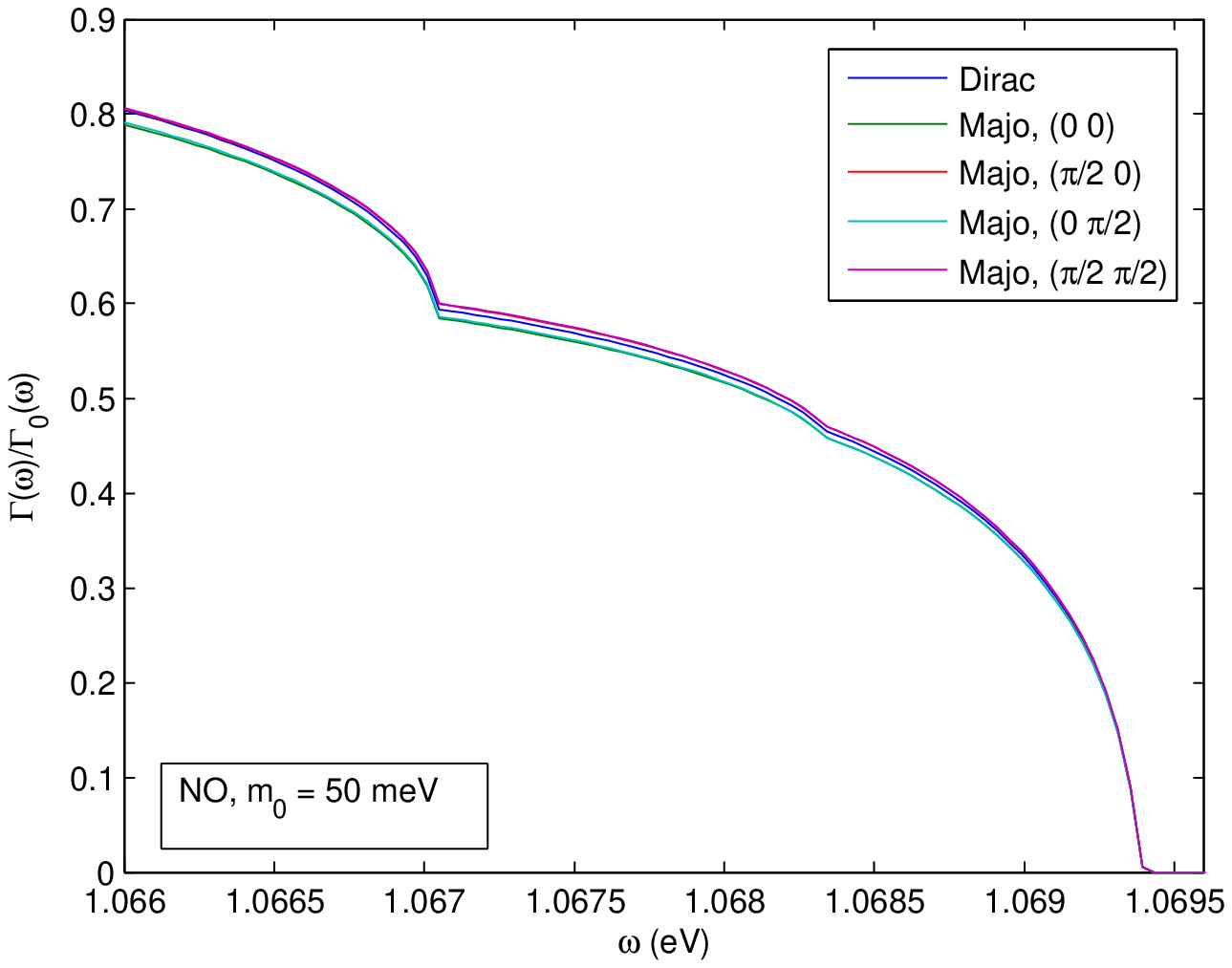} &
\includegraphics[width=7.5cm,height=6.5cm]{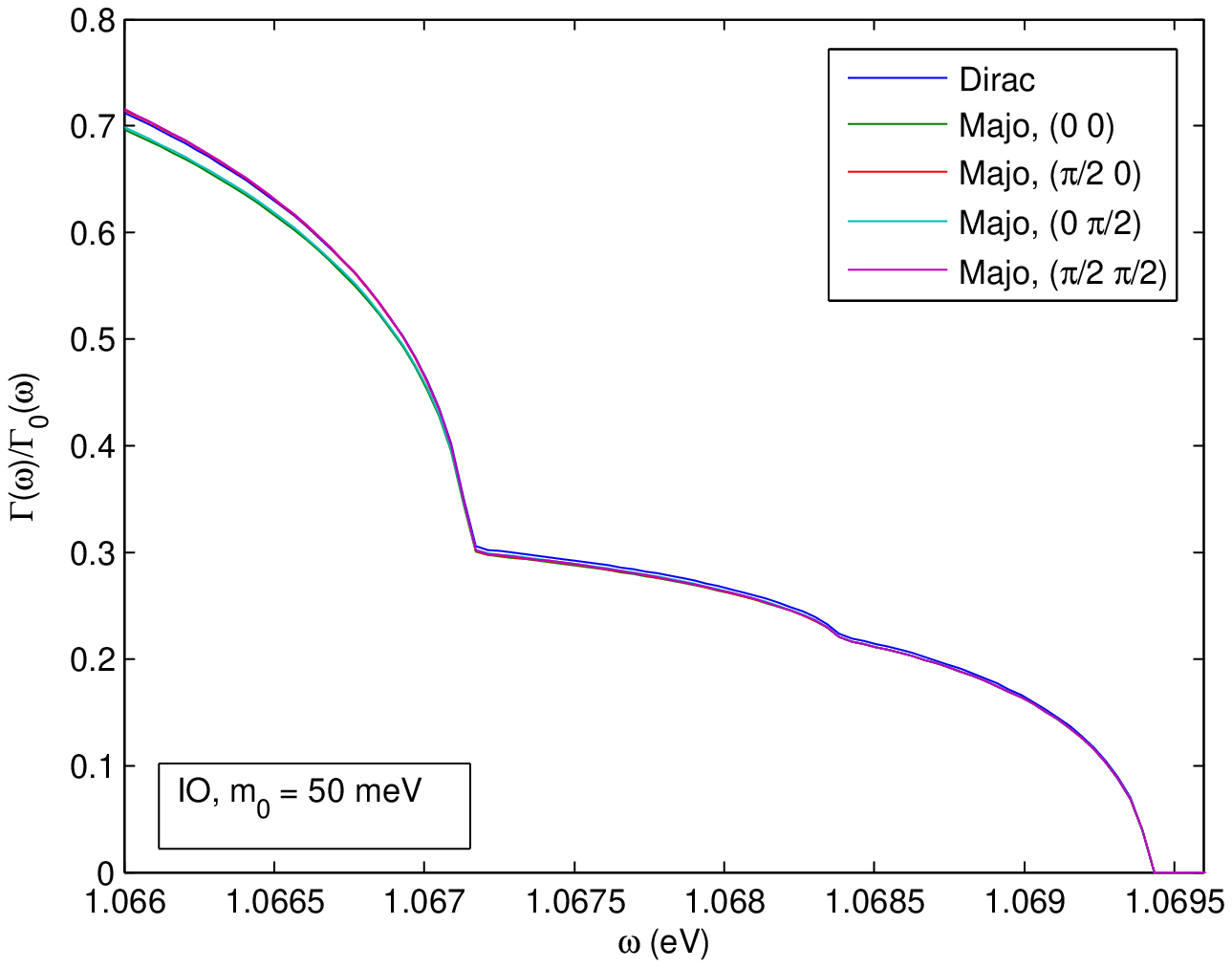}\\
\includegraphics[width=7.5cm,height=6.5cm]{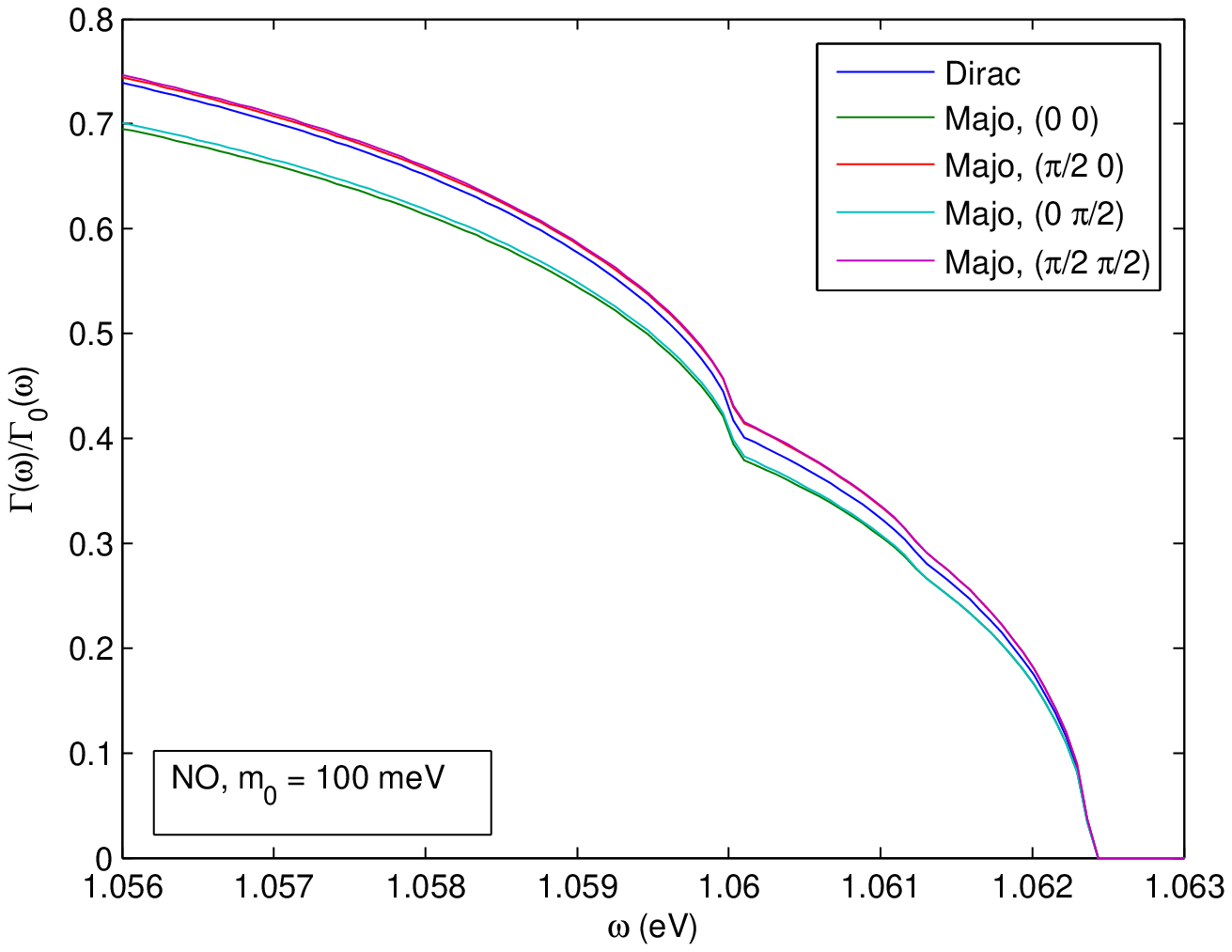} &
\includegraphics[width=7.5cm,height=6.5cm]{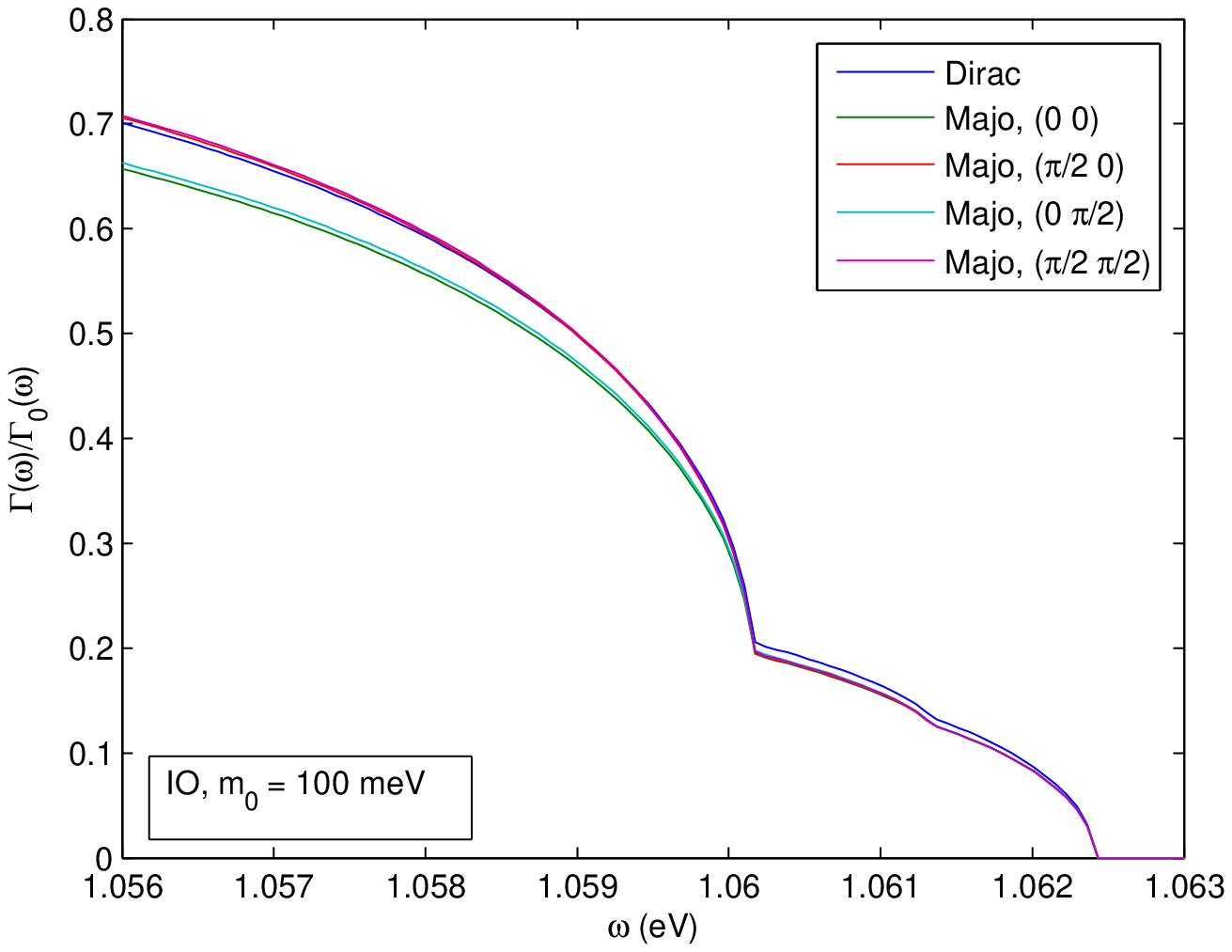}
\end{tabular}
\caption{The ratio
$R(\Gamma) \equiv
\Gamma_{\gamma 2\nu}(\omega)/\Gamma_{\gamma 2\nu}(\omega; m_i=0)
= I(\omega)/I(\omega; m_i=0)$
as a function of $\omega$ in the case of
emission of Dirac and Majorana
massive neutrinos having NO (left panels) or IO (right panels)
mass spectrum
corresponding to $m_0 = 50;100$ meV,
for  $\epsilon_{eg}=2.14$ eV and four values of
the CPV phases $(\alpha,\beta-\delta)$ in the Majorana case.
}
\label{QD_Eeg214}
\end{center}
\end{figure}
%
%
The latter does not corresponds to the features
shown in Fig. \ref{yb threshold}
(and in the subsequent figures of the paper),
where the kink due to the $\omega_{11}$
threshold is too small to be seen and
only the kink due to the $\omega_{12}$
threshold is prominent.\\

{\bf The Nature of Massive Neutrinos.}
 The Majorana vs Dirac neutrino distinction is much
more challenging experimentally,
if not impossible, with the Yb atom. This
is illustrated in Fig. \ref{Yb-Dirac-Majorana},
where the Dirac and Majorana spectra are almost
degenerate for both the NH and IH  cases.
The figure is obtained for $m_0 = 20$ meV and
the CPV phases set to zero, $(\alpha, \beta-\delta)=(0,0)$,
but the conclusion is valid for other choices of the
values of the phases as well.

 The difference between the emission of pairs of Dirac and Majorana
neutrinos can be noticeable in the case of QD spectrum with
$m_0\sim 100$ meV and for values of the phases $\alpha\cong 0$,
as is illustrated in Fig. \ref{QD_Eeg214},
where we show the ratio $R(\Gamma) \equiv
\Gamma_{\gamma 2\nu}(\omega)/\Gamma_{\gamma 2\nu}(\omega; m_i=0)
= I(\omega)/I(\omega; m_i=0)$ as a function of $\omega$.
As Fig. \ref{QD_Eeg214} indicates, the relative difference between
the Dirac and Majorana spectra can reach approximately 6\%
at values of $\omega$ sufficiently far below the threshold
energies $\omega_{ij}$. For $m_0 = 50$ meV, this difference
cannot exceed 2\% (Fig. \ref{QD_Eeg214}).

 A lower atomic energy scale $\epsilon_{eg} > 100$ meV,
which is closer in value to the largest neutrino mass,
would provide more favorable conditions for
determination of the nature of massive neutrinos and
possibly for getting information about at least some
(if not all) of the CPV phases.
In view of this we  now consider a
\begin{figure*}[t]
 \begin{center}
 \epsfxsize=0.6\textwidth
 \centerline{\epsfbox{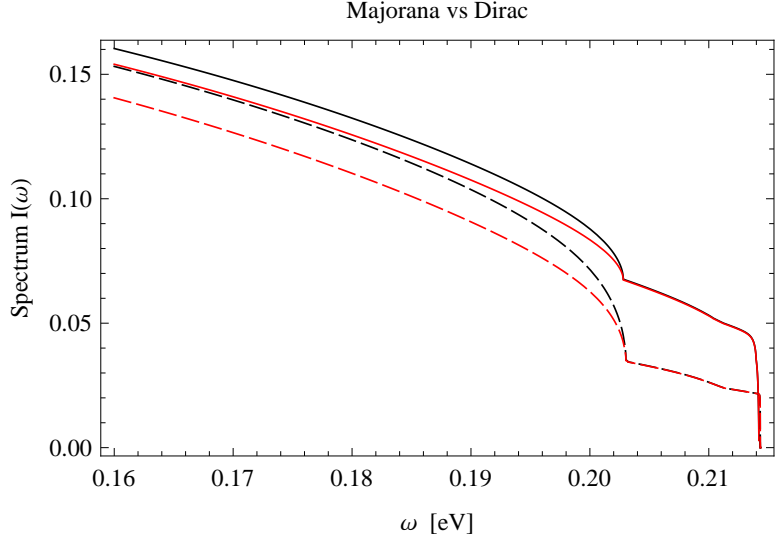}} \hspace*{\fill}
   \caption{Majorana vs Dirac neutrino comparison
in the case of X $^3P_0 \rightarrow {^1S_0}$ transitions with
energy difference $\epsilon_{eg}=\epsilon_{eg}({\rm Yb})/5$
for $m_{0}=2$ meV and NH  (solid lines) and IH (dashed lines) spectra.
The red and black lines correspond respectively to
Majorana and Dirac massive neutrinos.
}
   \label{MajoranavsDiracm0-2-Xfig04}
 \end{center}
\end{figure*}
%
hypothetical atom X scaled down in energy by 1/5
from the real Yb, thus $\epsilon_{eg} \sim 0.4$ eV.
There may or may not
be  good candidate atoms/molecules experimentally
accessible, having
level energy difference of order of the indicated value.
Figure \ref{MajoranavsDiracm0-2-Xfig04} shows
comparison between  spectra from X $^3P_0 \rightarrow ^1S_0$
for Majorana and Dirac neutrinos
with $m_{0}=2$ meV, for both the NH and IH cases.
As seen in Fig. \ref{MajoranavsDiracm0-2-Xfig04},
the Majorana vs Dirac difference is bigger than 5\% (10\%)
above the heaviest pair threshold in the NH (IH) case.
\begin{figure}[t]
\begin{center}
\begin{tabular}{cc}
\includegraphics[width=7.5cm,height=6.5cm]{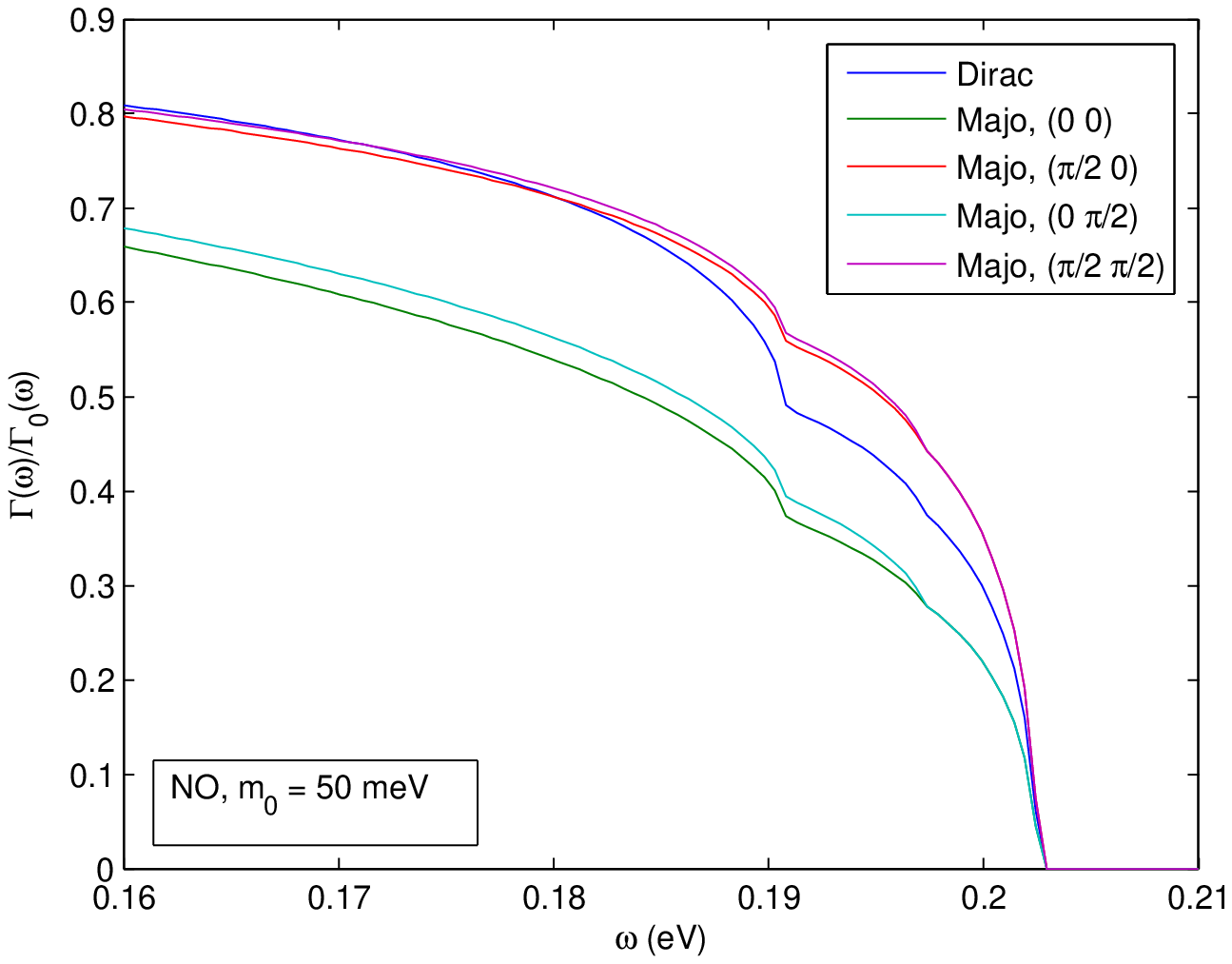} &
\includegraphics[width=7.5cm,height=6.5cm]{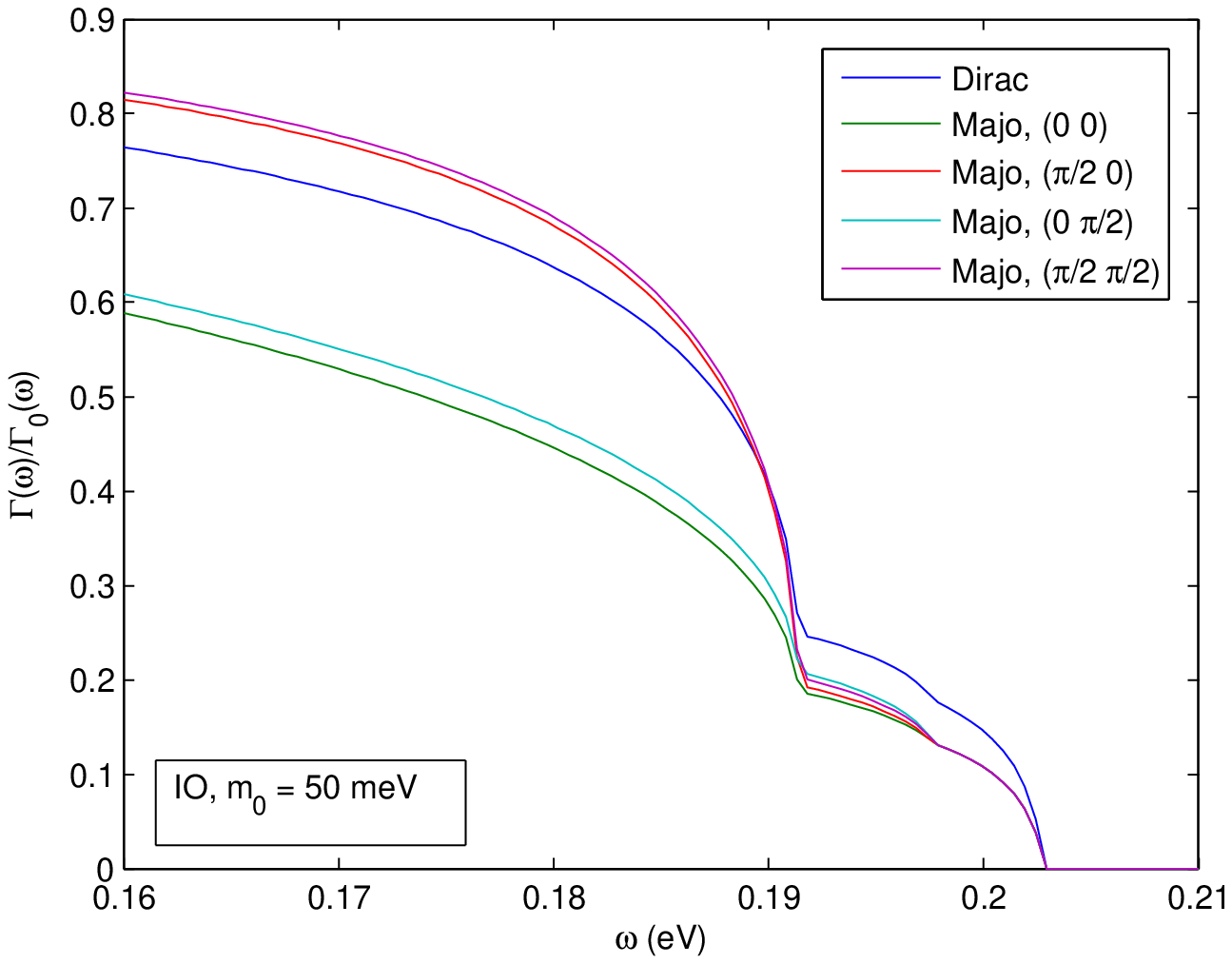}\\
\includegraphics[width=7.5cm,height=6.5cm]{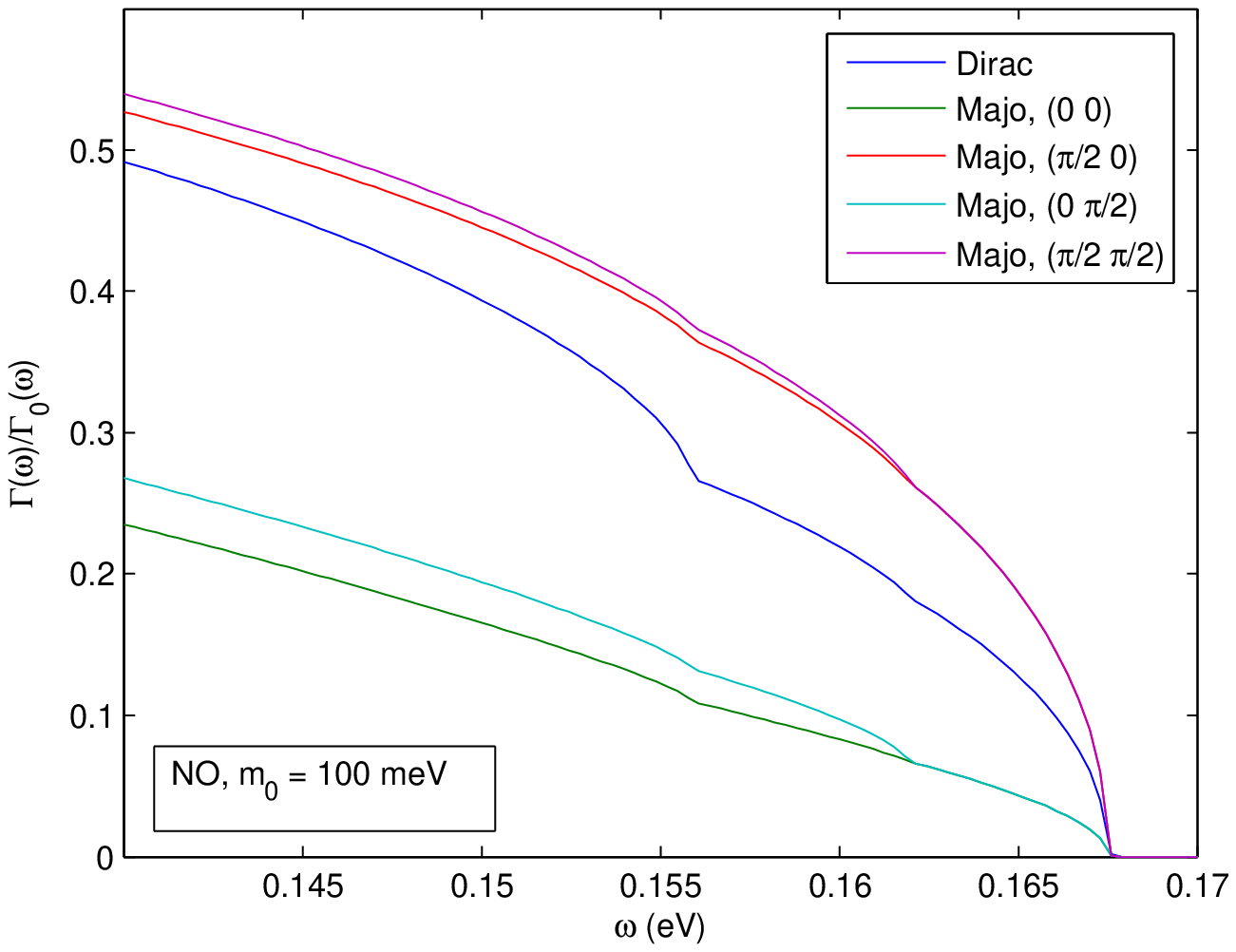} &
\includegraphics[width=7.5cm,height=6.5cm]{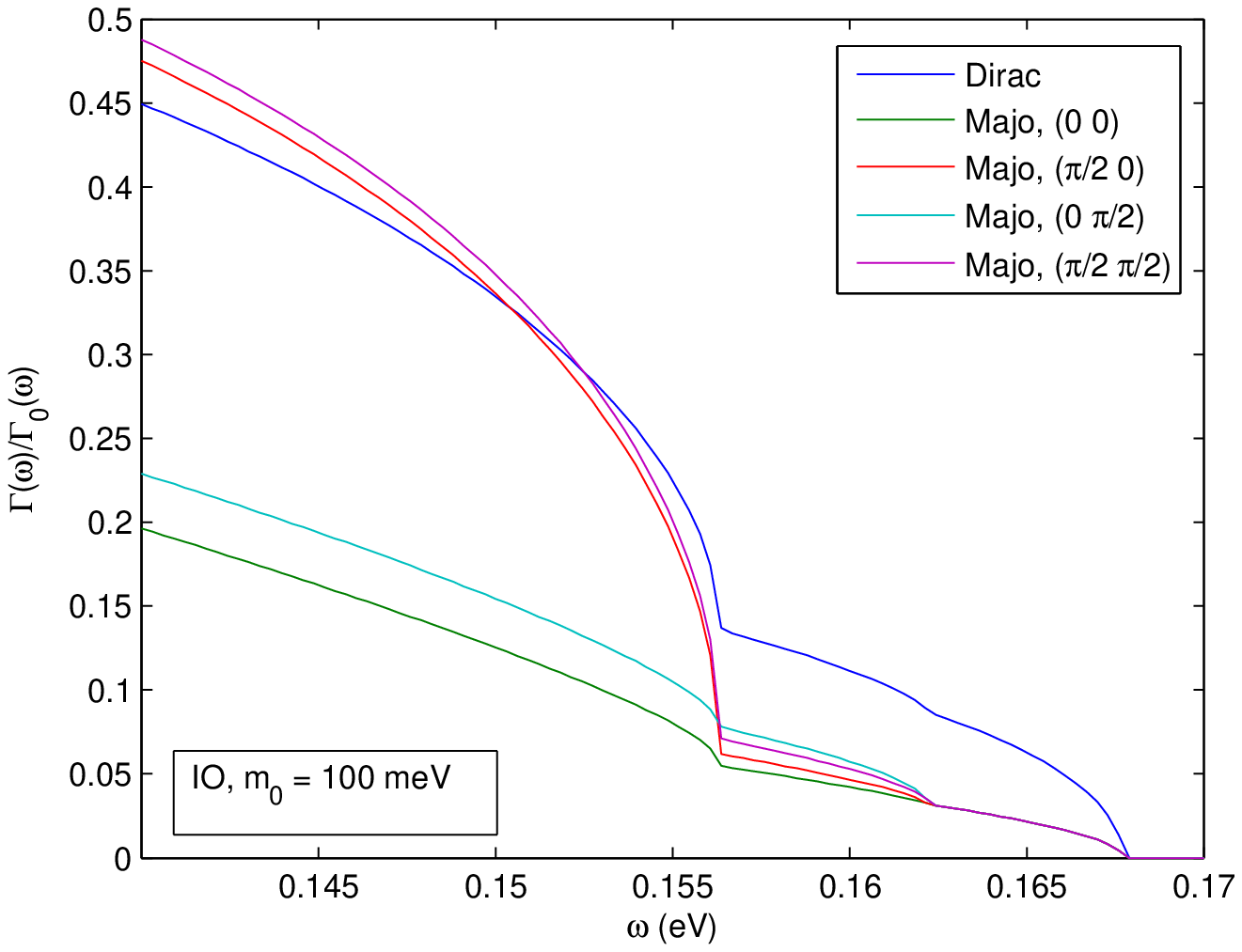}
\end{tabular}
\caption{The same as in Fig. \ref{QD_Eeg214}
but for $\epsilon_{eg}=0.43$ eV.
}
\label{QD_Eeg043}
\end{center}
\end{figure}
%
\begin{figure*}[htbp]
 \begin{center}
 \epsfxsize=0.6\textwidth
 \centerline{\epsfbox{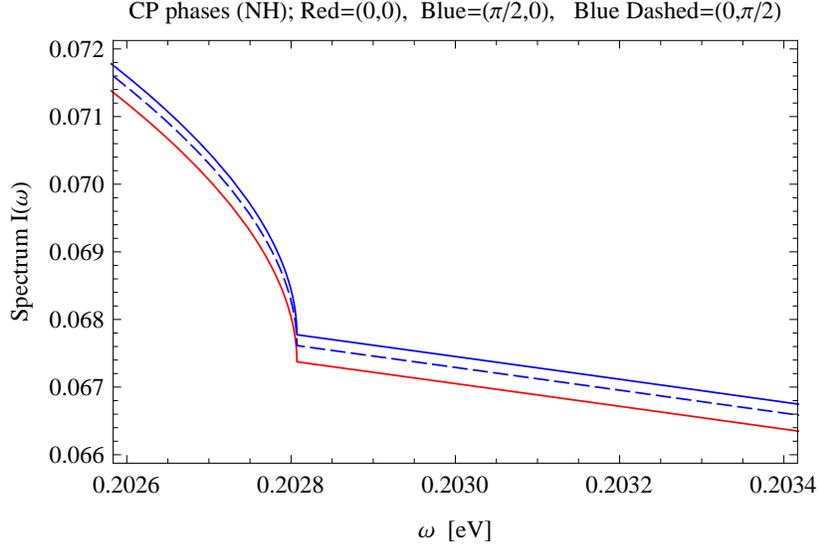}} \hspace*{\fill}
   \caption{
The dependence of $I(\omega)$ on the CPV phases $\alpha$ and
$(\beta - \delta)$ in the case of NH spectrum
with $m_0 = 2$ meV and for the transitions
corresponding to Fig. \ref{MajoranavsDiracm0-2-Xfig04}.
The red, solid blue and dashed blue lines
are obtained for $(\alpha,\beta-\delta) = (0,0),(\pi/2,0)$ and $(0,\pi/2)$,
respectively.
}
   \label{NormalCP-m0-2-Xfig12}
 \end{center}
\end{figure*}
%
\begin{figure}[htbp]
	\epsfxsize=0.6\textwidth
	\centerline{\epsfbox{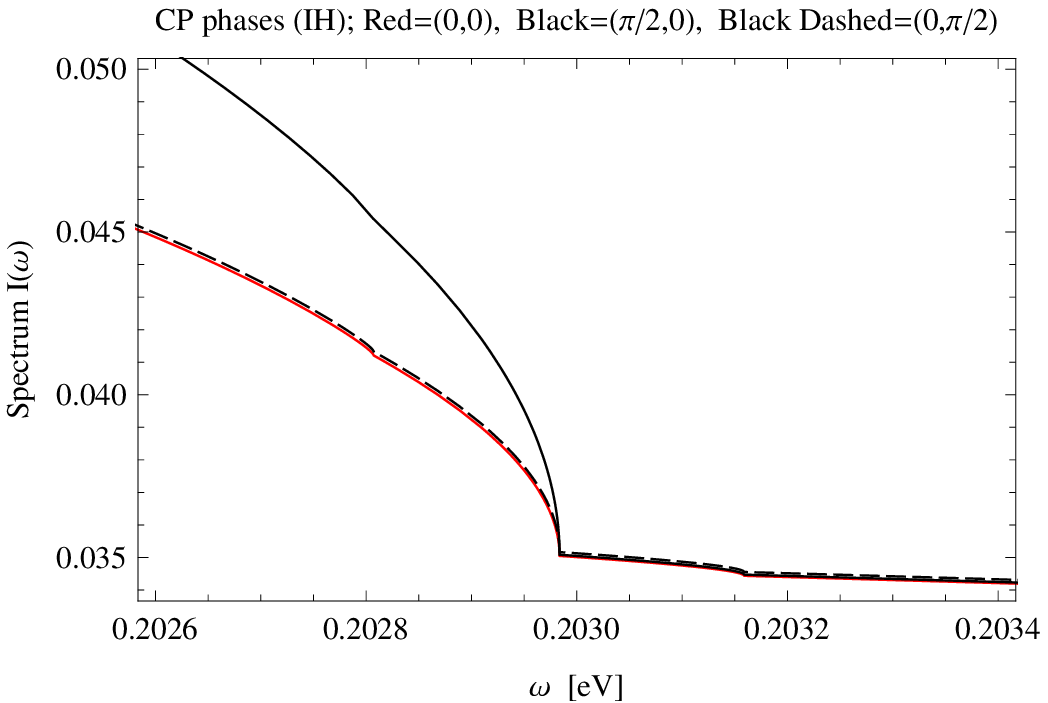}}
	\caption{The same as in Fig. \ref{NormalCP-m0-2-Xfig12}
for IH spectrum. The red, black solid and black dashed lines
correspond to $(\alpha,\beta-\delta) = (0,0),(\pi/2,0)$ and $(0,\pi/2)$,
respectively.
}
	\label{InvrtCP-m0-2-Xfig22}
\end{figure}
%
The difference becomes bigger for larger values of
the smallest neutrino mass $m_{0}$,
making the measurement easier.
This is illustrated in Fig. \ref{QD_Eeg043},
where we show again the ratio
$R(\Gamma) = I(\omega)/I(\omega; m_i=0)$ as a function of $\omega$
in the case of Dirac and Majorana pair neutrino emission
for $m_0 = 50;~100$ meV and NO and IO spectra. In the Majorana
neutrino case, the ratio $R(\Gamma)$ is plotted
for the four combinations of CP conserving values of the
phases $(\alpha,\beta-\delta) = (0,0);(0,\pi/2);(\pi/2,0);(\pi/2,\pi/2)$.
There is a significant difference between the
Majorana neutrino emission rates corresponding to
$(\alpha,\beta-\delta) = (0,0)$ and $(\pi/2,\pi/2)$.
The difference between the emission rates of Dirac and
Majorana neutrinos is largest for $(\alpha,\beta-\delta) = (0,0)$.
For $m_0 = 50~(100)$ meV and  $(\alpha,\beta-\delta) = (0,0)$.
for instance, the rate of emission of Dirac neutrinos
at $\omega$ sufficiently smaller than $\omega_{33}$ in the
NO case and  $\omega_{22}$ in the IO one,
can be larger than the rate of Majorana neutrino
emission by $\sim 20\%~(70\%)$.
The Dirac and Majorana neutrino
emission spectral rates never coincide.

 In Figs. \ref{NormalCP-m0-2-Xfig12} and
\ref{InvrtCP-m0-2-Xfig22} we show the spectral
rate dependence on the CPV phases $\alpha$ and $\beta -\delta$
for $m_0 = 2$ meV.
Generally speaking, the CPV phase measurement is
challenging, requiring a high statistics data acquisition.
A possible exception is the case of $\alpha$
and IH spectrum, as shown in Fig. \ref{InvrtCP-m0-2-Xfig22},
where the difference between the spectral rates
for $\alpha=0$  and $\alpha=\pi/2$ can reach  10\%.
For the NH spectrum, the analogous difference is at most
a few percent; observing this case requires large statistics
in actual measurements.

 It follows from these results  that one of
the most critical atomic physics parameters
for the potential of an RENP experiment
to provide information on the largest number of
fundamental neutrino physics observables of interest
is the value of the energy difference $\epsilon_{eg}$.
Values  $\epsilon_{eg} \leq $0.4 eV are favorable
for determining the nature of massive neutrinos,
and, if neutrinos are Majorana particles, for getting
information about at least some of the leptonic CPV phases,
which are the most difficult neutrino related observables
to probe experimentally.

\section{Summary and conclusion}

In the present work we investigated
the sensitivity to undetermined
neutrino parameters and properties
(the absolute mass scale, the type of neutrino mass spectrum,
the nature - Dirac or Majorana, of massive neutrinos
and the CP violating phases) of
the observables in macro-coherent RENP experiments.
The specific case of a potential RENP experiment measuring the
photon spectrum originating from  $^3P_0 \rightarrow {^1S_0}$
transitions in Yb atoms was considered. The relevant
atomic level energy difference is $\epsilon_{eg} = 2.14349$ eV.
Our results show that once the RENP events are unambiguously
identified experimentally, the least challenging would be
the measurement of the largest neutrino mass
(or the absolute neutrino mass scale).
The next in the order of increasing difficulty is the
determination of the neutrino mass spectrum or hierarchy (NH, IH, QD).
The Majorana vs Dirac distinction and the measurement of
the CPV phases are considerably more challenging,
requiring high statistics data from
atoms (or molecules) with lower  energy
difference $\epsilon_{eg} \ltap 0.5$ eV.
Although the measurements of the indicated
fundamental parameters of neutrino physics
might be demanding, a single RENP experiment
might provide a systematic strategy to determine
almost all of these parameters, and thus can contribute
to the progress in understanding
the origin of neutrino masses
and of the
physics beyond the Standard Model
possibly associated with their existence.

 The present work points to the
best atom/molecule candidate
with level energy difference of
less than  O(0.5 eV) for the indicator
$\epsilon_{eg}$. Besides the desirable richness of
detectable observables, good candidates for realistic
RENP experiments have to be searched also from the point
of least complexity of target preparation.
Investigations along these lines are in progress
by a group including some of us.

\vspace{1cm}
{\bf Acknowledgements}

Two of us (N.S. and M.Y.) are grateful to S. Uetake for a discussion
on Yb atomic data. The research of N.S., M.T., and M.Y.
was partially supported by Grant-in-Aid for Scientific
Research on Innovative Areas "Extreme quantum world opened up by atoms"
(21104002) from the Ministry of Education,
Culture, Sports, Science, and Technology of Japan.
This work was supported in part by the INFN program on
``Astroparticle Physics'', by the Italian MIUR program on
``Neutrinos, Dark Matter and  Dark Energy in the Era of LHC''
(D.N.D. and S.T.P.) and by the World Premier International
Research Center Initiative (WPI Initiative), MEXT, Japan  (S.T.P.).

\end{document}